\documentclass[
twocolumn,  
secnumarabic,
amssymb, 
nobibnotes,
aps, 
prd,
superscriptaddress,
nofootinbib,
]{revtex4-1}
\usepackage{algorithm}
\usepackage{placeins}

\newcommand{\diff}{{\rm d}}

\usepackage[ddmmyyyy]{datetime} 
\usepackage{amsmath} 
\usepackage[dvipsnames]{xcolor}
\usepackage{cancel} 
\usepackage{graphicx}
\usepackage{comment} 
\usepackage{soul,ulem} 
\usepackage{eqnarray}
\usepackage{multirow}
\soulregister\cite7
\soulregister\ref7 
\soulregister\eqref7 
\usepackage{bbding}
\usepackage{pifont}
\usepackage{wasysym}
\usepackage{amssymb}
\usepackage{textcomp} 
\usepackage{marginnote} 
\usepackage[colorinlistoftodos
, textsize=tiny]{todonotes} 
\usepackage{textcomp} 
\usepackage{lineno} 
\usepackage{hyperref}
\usepackage{cleveref}
\hypersetup{
    colorlinks=true,
    citecolor= teal, 
    linkcolor= [rgb]{0.1,0.5,0.5},
    filecolor=magenta,      
    urlcolor=[rgb]{0.1,0.5,0.5},
    urlbordercolor=red
}

\newcommand*\aap{Astron.~Astrophys.}

\newcommand*\aapr{Astron.~Astrophys.~Rev.}

\newcommand*\aj{AJ}
\newcommand*\apjl{Astrophys.~J.~Lett}

\newcommand*\jcap{J. Cosmology Astropart. Phys.}

\newcommand*\mnras{Mon.~Not.~Roy.~Astron.~Soc.}

\newcommand{\GNeff}{\ensuremath{G_{\rm N}^{\rm eff}}}
\newcommand{\gNtilde}{\ensuremath{\tilde{\gamma}_{\rm N}}}

\newcommand{\GN}{\ensuremath{G_{\rm N}}}

\newcommand{\Mhs}{\ensuremath{M_{200}^{\rm{HS}}}}
\newcommand{\Rhs}{\ensuremath{R_{200}^{\rm{HS}}}}
\newcommand{\Mcau}{\ensuremath{M_{200}^{\rm{Cau}}}}
\newcommand{\Rcau}{\ensuremath{R_{200}^{\rm{Cau}}}}
\newcommand{\Mjoint}{\ensuremath{M_{200}^{\rm{Joint}}}}
\newcommand{\Munit}{\ensuremath{10^{14}\, M_{\odot}}}

\newcommand{\M}{\ensuremath{M_{500}}}

\newcommand{\Md}{M_{\Delta}}
\newcommand{\cd}{c_{\Delta}}
\newcommand{\Rd}{R_{\Delta}}

\newcommand{\ksM}{\text{km/s Mpc$^{-1} $}}

\newcommand{\fbr}{\ensuremath{\mathcal{F}_{\beta}(r)}}
\newcommand{\Mpc}{{\rm Mpc})}

\newcommand{\phitwo}{\ensuremath{\phi_{\infty, 2}}}
\newcommand{\btwo}{\ensuremath{\beta_{2}}}
\newcommand{\LCDM}{\text{$\Lambda$CDM} }

\newcommand{\fofr}{\text{$f(R)$} }

\defcitealias{Ettori:2018tus}{E19}

\begin{document}

\title{ Caustic and hydrostatic mass bias: Implications for modified gravity}

\author{Minahil Adil Butt}
\email{mbutt@sissa.it}
\affiliation{SISSA-International School for Advanced Studies, Via Bonomea 265, 34136 Trieste, Italy}
\affiliation{ICTP-The Abdus Salam International Centre for Theoretical Physics, Strada Costiera 11, 34151 Trieste, Italy  }
\affiliation{IFPU, Institute for Fundamental Physics of the Universe, via Beirut 2, 34151 Trieste, Italy}
\affiliation{INFN, Sezione di Trieste, Via Valerio 2, I-34127 Trieste, Italy}

\author{Balakrishna S. Haridasu}
 \email{sharidas@sissa.it}
 \affiliation{SISSA-International School for Advanced Studies, Via Bonomea 265, 34136 Trieste, Italy}
 \affiliation{IFPU, Institute for Fundamental Physics of the Universe, via Beirut 2, 34151 Trieste, Italy}
 \affiliation{INFN, Sezione di Trieste, Via Valerio 2, I-34127 Trieste, Italy}

\author{Yacer Boumechta}
\affiliation{SISSA-International School for Advanced Studies, Via Bonomea 265, 34136 Trieste, Italy}
\affiliation{ICTP-The Abdus Salam International Centre for Theoretical Physics, Strada Costiera 11, 34151 Trieste, Italy  }
\affiliation{IFPU, Institute for Fundamental Physics of the Universe, via Beirut 2, 34151 Trieste, Italy}
\affiliation{INFN, Sezione di Trieste, Via Valerio 2, I-34127 Trieste, Italy}
 
\author{Francesco Benetti}
\affiliation{SISSA-International School for Advanced Studies, Via Bonomea 265, 34136 Trieste, Italy}

 \author{Lorenzo Pizzuti}
 \email{pizzuti@oavda.it}
 \affiliation{Dipartimento di Fisica G. Occhialini, Universit\'a degli Studi di Milano Bicocca, Piazza della Scienza 3, I-20126 Milano, Italy}

\author{Carlo Baccigalupi}
\email{bacci@sissa.it}
\affiliation{SISSA-International School for Advanced Studies, Via Bonomea 265, 34136 Trieste, Italy}
 \affiliation{IFPU, Institute for Fundamental Physics of the Universe, via Beirut 2, 34151 Trieste, Italy}
 \affiliation{INFN, Sezione di Trieste, Via Valerio 2, I-34127 Trieste, Italy}

\author{Andrea Lapi}
\email{lapi@sissa.it}
\affiliation{SISSA-International School for Advanced Studies, Via Bonomea 265, 34136 Trieste, Italy}
 \affiliation{IFPU, Institute for Fundamental Physics of the Universe, via Beirut 2, 34151 Trieste, Italy}
 \affiliation{INFN, Sezione di Trieste, Via Valerio 2, I-34127 Trieste, Italy}
 \affiliation{IRA-INAF, Via Gobetti 101, 40129 Bologna, Italy}

\begin{abstract}

We propose and perform a joint analysis of the two different mass estimates of galaxy clusters, namely the hydrostatic and caustic techniques. Firstly, we show comprehensively that the mass bias between these two techniques can be possibly alleviated when cluster-specific assumptions constrained using the hydrostatic technique are utilized within the caustic technique. While at face value this demotes the caustic technique from a completely independent method, this allows one to further tighten the constraints on the cluster mass and subsequently, allow us to test modifications to gravity. Implementing the aforementioned formalism for two well-observed massive galaxy clusters, A2029 and A2142, we highlight the proof of concept. In the current implementation, we use this method to constrain the Chameleon screening and Vainshtein screening. As anticipated, we show that the joint analysis can help improve the constraints on these modified gravity scenarios. 

\end{abstract}

\maketitle


\section{Introduction}
\label{sec:Introduction}

The accurate determination of galaxy cluster masses stands as a cornerstone in our quest to understand the evolution of cosmic structures and exploit clusters \cite{Eckert_2022, Aarseth_1979, de_Andres_2022} as indispensable cosmological tools \cite{Cui_2018, Planck:2015fie, Planck:2015mrs}. Beyond their significance in the broader cosmological context, cluster mass determination holds a pivotal role in deciphering the intricate astrophysical processes governing both baryonic and dark matter physics within these clusters, encompassing the hot gas of the intracluster medium (ICM) and member galaxies, as well as the thermodynamical properties of the clusters in the outskirts \cite{Eckert:2013faa, Eckert:2015rlr, Ettori:2013tka, Ettori:2018tus}.\\

Cluster mass measurement, however, presents an intricate challenge, primarily because the dominant dark matter component can only be indirectly probed, often relying on assumed fitting functions such as the Navarro-Frenk-White (NFW) profile \cite{Navarro:1995iw}. Determining the total mass of a cluster traditionally hinges upon the gravitational influence it exerts on the properties of the ICM and member galaxies \cite{Ettori_2019}, or its gravitational lensing effect on the light originating from background sources \cite{Murray_2022}.
The most accurate and precise mass estimation techniques include hydrostatic masses determined from X-ray observations of the ICM, caustic techniques based on galaxy dynamics \cite{Diaferio99}, kinematics of the galaxy cluster members solving the Jeans equation (for example \cite{Biviano23}) and weak gravitational lensing measurements \cite{von_der_Linden_2014}. 

The gold standard for cluster mass estimation has long been hydrostatic masses derived from X-ray observations of the ICM \cite{Ettori_2019}. Nevertheless, the accuracy of these estimates can be compromised by departures from hydrostatic equilibrium or the presence of non-thermal pressure sources, including turbulence and cosmic rays, often leading to systematic underestimations of the true mass by 10-30 percent \cite{Ettori_2019, Hurier_2018}. Recent advances in both techniques and data quality have spotlighted these biases, sparking a reevaluation of hydrostatic mass estimates against alternative methods such as weak gravitational lensing \cite{Maughan_2016, Ettori_2019}.
Among the most robust mass estimation techniques, weak gravitational lensing measurements, which probe the distortion of background source light due to the cluster's gravitational field, offer unparalleled precision and independence from dynamical assumptions. Nevertheless, they too necessitate high-quality data, including extensive galaxy redshifts and lensed sources{, as well as assumptions in de-projection to reconstruct the shape of the three-dimensional mass profile (e.g. \cite{Umetsu2020}).}\\
This paper embarks on an exploration of caustic techniques for mass profiling, serving as a promising complement to weak gravitational lensing \cite{Amon_2018, Tihhonova_2018}. The caustic method relies on identifying unique structures within line-of-sight velocity and projected-radius space that mirror the cluster's escape velocity profile. These caustics provide a means to reconstruct the cluster's enclosed mass, extending far beyond the virial radius \cite{Diaferio99, Gifford13, Serra_2010}.\\
Similar to the lensing measurements \cite{pan2023measurement, High_2012,Diaferio_2005}, caustic masses remain agnostic to the dynamical state of clusters and are resilient against the physical processes that can bias hydrostatic mass estimates. In practice, the caustic technique capitalizes on overdense envelopes in phase-space diagrams, mapping the line-of-sight velocity versus projected radius for galaxies inhabiting cluster infall regions \cite{Diaferio_1997, Diaferio99, Diaferio09}. These envelopes' edges, referred to as the caustic profile ($\mathcal{A}(r)$), can be extracted from observational data \cite{Serra_2010}, though needing intricate techniques. An inherent assumption in caustic mass estimates is the constancy of the filling factor $\mathcal{F}_{\beta}(r)$, a parameter linked to the ratio of mass gradient to gravitational potential \cite{Diaferio99, Serra_2010}. Nevertheless, N-body simulations have unveiled breakdowns in this approximation within cluster inner regions, potentially resulting in an overestimate of up to 10-20 percent at $r_{500}$, {i.e. the radius enclosing an average overdensity of $500$ times the critical density of the universe at that redshift (see Section \ref{sec:Analysis})}, increasing at smaller radii \cite{Serra_2010}. While hydrostatic and caustic methods operate independently, each resting on distinct assumptions and systematic uncertainties, they exhibit reasonable agreement on average \cite{Logan_2022, Andreon_2017}. While a significant scatter is present between the two estimators, mass estimates align to within approximately 20 percent over the full radial range sampled by both techniques \cite{Maughan_2016}. 

In the pursuit of alleviating biases between hydrostatic and caustic masses \cite{Maughan_2016, Logan_2022}, we speculate on the effects of relaxing the constant $\mathcal{F}_{\beta}(r)$ assumption \cite{Diaferio99, Gifford13}. To this end, we explore varying $\mathcal{F}_{\beta}(r)$ under the framework of General Relativity (GR) (assuming a $\Lambda$CDM Universe), and two modified gravity models: Chameleon \cite{Terukina:2013eqa} and Vainshtein Screening \cite{Babichev_2013}. The Lagrangian of the Chameleon screening theory includes the usual Einstein-Hilbert Lagrangian plus the scalar field, in addition to the Standard Model fields coupled minimally to gravity
\cite{Khoury:2003aq,Zaregonbadi:2022lpw,Ivanov:2016ucz,Kraiselburd:2015vyf,Tsujikawa:2009yf}. 
In this model we obtain interesting phenomenology on cosmological scales while simultaneously screening MG effects in high-density environments, allowing it to avoid solar system constraints. Thus, GR is obtained in regions of high-density while in low-density environments we observe the effects of the modified potentials. The Vainshtein screening \cite{Vainshtein:1972sx} is a
useful mechanism for the higher-order scalar tensor \cite{Babichev_2013,Brax:2004qh}. In Vainshtein screening, the gravitational potentials are modified inside the matter sources \cite{Crisostomi_2018,Langlois_2018,Bartolo_2018,Dima_2018,Hirano:2019nkz,Crisostomi:2019yfo}. This mechanism allows to hide via non-linear effects - typically for source distances smaller than a so-called Vainshtein radius which depends on the source and the theory considered - some degrees of freedom whose effects are then only left important at large distances, e.g. for cosmology. Both these theories have been tested using galaxy clusters \cite{Terukina14, Haridasu:2021hzq}.

Our analysis aims to shed light on the resultant impact on the caustic mass and, by extension, the mass bias, scrutinizing diverse regions of the modified gravity parameter space. The structure of this paper unfolds as follows: \Cref{sec:Caustic} offers a concise overview of the Caustic technique. In \Cref{sec:screening}, we delve into the intricacies of screening mechanisms within the framework of modified gravity scenarios. \Cref{sec:Analysis} outlines our approach, encompassing data sources, methodology, and analytical techniques. Our results, from the current investigation, are presented in \Cref{sec:results}. Finally, \Cref{sec:conclusion} provides concluding remarks. Throughout this work, we assume $H_0 = 70 \, \ksM$ and $\Omega_m = 0.3$ in the estimation of the critical density $\rho_{\rm{crit}} = 8 \pi G/ 3 H(z)^2 $ with the standard $\LCDM$ background having $H(z)^2 = \Omega_m (1+z)^3 + 1- \Omega_m$.


\section{The Caustic Method}
\label{sec:Caustic}
The caustic technique, as introduced by \cite{Diaferio_1997} and further developed in subsequent works (\cite{Diaferio99, Diaferio09, Serra_2010}), provides a unique approach to estimate the mass of the clusters, utilizing the escape velocity profile of member galaxies within the cluster. This technique extends our insights from the central cluster region to radii as large as three times the cluster's virial radius, denoted as $r_{200}$, where $r_{200}$ signifies the radius of a sphere with an average density 200 times the critical density ($\rho_{\rm{crit}}$) of the background universe. This is particularly valuable because, at such large radii, galaxy clusters might not be in complete dynamical equilibrium, based solely on galaxy redshift data.\\
Hierarchical clustering, a prominent formation mechanism for galaxy clusters, involves the aggregation of smaller systems. Unlike a purely radial infall expected in the spherical collapse model, this process incorporates substantial non-radial velocities (e.g., White et al. 2010), resulting in a complex velocity distribution of the galaxies within clusters (see reviews in \cite{Diaferio09} and \cite{Serra_2010}). The primary reason for this distribution is completely dependent on the local gravitational potential \cite{Diaferio_1997}, coupled with the influences of surrounding groups and tidal fields. When visualized in a redshift diagram, portraying the line-of-sight velocity versus projected distance from the cluster center {(the so-called projected phase space, p.p.s.)}, cluster members delineate a distinctive trumpet-shaped region symmetric along the radial axis \cite{1987MNRAS.227....1K, 1993ApJ...418..544V, 1989AJ.....98..755R}.\\
Within this framework, the caustic surface serves as a crucial demarcation, defining the boundaries of this trumpet-shaped region. The amplitude of these caustics, denoted as $\mathcal{A}(r)$, diminishes as we move away from the cluster center, and it is intrinsically tied to the average velocity component $\langle v^{2}\rangle$ \cite{Diaferio_2005}. In a spherically symmetric system, the escape velocity $v_{\rm{esc}}^{2}(r)$ is directly related to the gravitational potential $\Phi(r)$ and is a non-increasing function of radial distance. The highest observable velocity corresponds to the escape velocity. Therefore, $\mathcal{A}^2(r)$ at a projected radius $r_{\perp}$ effectively measures the average velocity component along the line of sight at the three-dimensional radius $r = r_{\perp}$. 

To quantify this average velocity component, the velocity anisotropy profile $\beta(r)$ is utilized, which characterizes the velocity distribution's deviation from isotropy. This profile $\beta(r)$ is expressed as $1-(\langle v_{\theta}^{2}\rangle+ \langle v_{\phi}^{2}\rangle)/2\langle v_{r}^{2}\rangle$, where $v_{\theta}$, $v_{\phi}$, and $v_{r}$ are the longitudinal, azimuthal, and radial velocity components of galaxies, respectively (e.g. \cite{Diaferio99}). {The velocity anisotropy is generally an unknown variable in kinematics analyses, and it should be correctly modeled along with the other dynamical quantities. Parametric methods (e.g.  \cite{Mamon01,Read21,Biviano23}) assume specific - physically motivated - profiles which are fitted to the observed velocity and position fields. Kinematics determinations based on the Jeans' equation lead to a degeneracy between $\beta(r)$ and the total cluster mass (the so-called \textit{mass-anisotropy degeneracy,} MAD); if additional information on the mass profiles is provided (such as from X-ray or Weak Lensing analyses),  non-parametric reconstructions can be performed by inversion of the data (e.g. \cite{Binney1982}, see also \cite{Mamon19} and references therein). } 

The gravitational potential profile is related to the caustic amplitude through the function $g(\beta)$, given by:
\begin{equation}
    g(\beta)=\frac{3-2\beta(r)}{1-\beta(r)}.
    \label{eqn:gbeta}
\end{equation}
As such, after estimating the caustic amplitude, $\beta(r)$ becomes the sole unknown factor in estimating the gravitational potential. It's worth emphasizing that the caustic technique doesn't depend on assumptions regarding dynamical equilibrium, the shape of $g(\beta)$, or the gravitational potential profile $\Phi(r)$ individually. Instead, it quantifies the combined effect of $g(\beta)$ and $\Phi(r)$ in terms of the caustic amplitude $\mathcal{A}^2(r)$,   
\begin{equation}
    \label{eqn:phi_caustic}
        -2\Phi(r)=\mathcal{A}^{2}(r)g(\beta).
    \end{equation}
This equation implies that the caustic technique can estimate a combination of the gravitational potential profile and the velocity anisotropy parameter $\beta$. This provides valuable insights into the dynamical properties of spherical systems, particularly in the context of galaxy clusters and their mass profiles.\\
The equation for the caustic mass profile of a spherical system can be expressed as:
\begin{equation}
        \label{eqn:Mass_Profile}
        GM(<r)=\int_{0}^{r}\mathcal{A}^{2}(r)\mathcal{F}_{\beta}(r)dr
\end{equation}
where $\mathcal{F}_{\beta}(r) =\mathcal{F}(r)g(\beta)$ and 
\begin{equation}
    \label{eqn:fofr}
    \mathcal{F}(r)=-2\pi G\frac{\rho(r)r^{2}}{\Phi(r)}.
\end{equation}
In this context, $\rho(r)$ is the density profile of the spherical system (which we assume to be the NFW profile) and $\Phi(r)$ stands for the gravitational potential profile. \\
\Cref{eqn:Mass_Profile} relates the mass profile to the density profile of a spherical system and a profile cannot be inferred without knowing the other.
In hierarchical clustering scenarios, the function $\mathcal{F}(r)$ doesn't exhibit strong variations with respect to radial distance $r$ \cite{Diaferio_1997, Diaferio99}. Similarly, $\mathcal{F}_{\beta}(r)$ also changes slowly with $r$ if the velocity anisotropy parameter $\beta$ is governed by a slowly varying function $g(\beta)$.\\
{The practical utility of this technique becomes evident when applied to individual clusters. It typically provides accurate escape velocity and mass profiles, though there may be occasional deviations from the actual profile. For example, when the observed galaxy sample has low completeness\footnote{Completeness measures the number of member galaxies available in the sample with accurate spectroscopic redshifts \cite{Sohn_2017}. } towards the outskirts of the cluster, the estimated caustic surface could be subsequently lower and eventually provide a lower mass estimate. This is also very evident in the flattening of the mass profile when no galaxies are present within the sample (see also \Cref{sec:mass_profiles}).} This method holds significant relevance as an alternative to gravitational lensing for measuring mass in a cluster's outer regions \cite{Diaferio_2005}. Unlike lensing, it can be applied to clusters at any redshift, provided there are a sufficient number of galaxies for a proper redshift diagram analysis.\\
An important milestone in the application of the caustic technique was the work of Geller, Diaferio, and Kurtz in 1999 \cite{Geller:1999ci}. They employed this method to measure the mass profile of the Coma cluster, extending their analysis to an impressive $10 h^{-1}$Mpc from the cluster center. Their findings demonstrated that the Navarro, Frenk, \& White (NFW) profile provides an excellent fit to the cluster density profile at these extensive radii, thereby challenging the viability of the isothermal sphere as a cluster model. Subsequently, Biviano and Girardi (2003) \cite{Biviano_2003} applied the caustic technique to a composite cluster, stacking data from 43 clusters in the Two Degree Galaxy Redshift Survey (2dGFRS, Colless et al. 2001). Their results also aligned with earlier lensing and X-ray analyses, affirming the method's robustness and accuracy in estimating cluster mass profiles.

\section{Modified gravity and Screening Mechanisms}
\label{sec:screening}
In this section, we briefly describe the modified gravity scenarios assessed in the current work and the underlying screening mechanisms. 

\textit{Chameleon Screening}: The chameleon model \cite{Khoury:2003aq, Zaregonbadi:2022lpw, Tsujikawa:2009yf, Kraiselburd:2015vyf, PhysRevLett.93.171104} modifies gravity by introducing a scalar field non-minimally coupled with the matter components and gives rise to a fifth force that can be of the same order as the standard gravitational force. The chameleon mechanism is a two-parameter model. The chameleon model parameter $\beta$ determines the strength of the fifth force when it is not screened. The second chameleon parameter, $\phi_{\infty}$, controls the effectiveness of the screening mechanism, describing the transition from the inner region of a cluster where gravity may be Newtonian to the outer region where the fifth force contributes. The critical radius, where the transition occurs, is determined by both $\phi_{\infty}$ and $\beta$. In the absence of environmental effects, $\phi_{\infty}$ can be regarded to be the cosmological background
value of the chameleon field. 

The chameleon mechanism operates whenever a scalar field couples to matter in such a way that its effective mass depends on the local matter density.
The scalar-mediated force between matter
particles can be of gravitational strength, but its range is a decreasing function of ambient matter density, thereby avoiding detection in regions of high density. Deep in space, where the mass density is low, the scalar is light and mediates a fifth force of gravitational strength, but near the Earth, where experiments are performed, and where the local density is high, it acquires a large mass, making its effects short range and hence, unobservable. {This is achieved with a canonical scalar field with suitable self-interaction potential $V(\phi)$, which is a decreasing function of $\phi$ \cite{PhysRevLett.93.171104}. For instance, we assume a usual power-law potential of the form $V(\phi) = \Lambda^{n+4}\phi^{-n}$, however, the scalar field constrained within the cluster is not sensitive to parameters $\{\Lambda,\, n\}$ \cite{Terukina:2013eqa}.} The theory (in the weak-field limit and for the non-relativistic matter) is given by \cite{Terukina:2013eqa, khoury2013les}:
\begin{equation}
    L_{\mathrm{chameleon}}=-\frac{1}{2}(\partial\phi)^{2}-V(\phi)-\frac{g\phi}{M_{\mathrm{Pl}}}\rho_{\mathrm{m}}
\end{equation}
The dimensionless coupling parameter $g$ is assumed to be $O(1)$, corresponding to the gravitational strength coupling. In regions of high density, the mass of the chameleon field increases. The range of interaction decreases, thus screening the effect of the fifth force and recovering GR. In low-density environments, the fifth force is unscreened and the effects of the chameleon field can be observed. 
The modified gravitational potential under this mechanism is given by \cite{Terukina:2013eqa, Wilcox:2015kna}:

\begin{equation}
    \label{eqn:grav_cham}
    \frac{ \diff { \Phi(r)}}{\diff {r}}=\frac{G_{\mathrm{N}}{M}(r)}{r^{2}} + \beta \frac{\diff{\phi}}{\diff {r}}
\end{equation}
where $\phi(r)$ is the chameleon field.\\
The chameleon field mediates a long-range fifth force when the matter density is still large
compared to the background, and the scalar field has not settled in the minimum of the effective potential. The chameleon field becomes effective beyond a critical radius, $r_{c}$, below which, the field is completely screened. This radius is determined by \cite{Terukina:2013eqa}:
\begin{equation}
    \label{eqn:critical_radius}
    1+\frac{r_{c}}{r_{s}}=\frac{\beta\rho_{s}r_{s}^{2}}{M_{\mathrm{Pl}}\phi_{\infty}}.
\end{equation}
The chameleon field \cite{Terukina:2013eqa} in the two limits is given by:
\begin{equation}
    \label{eqn:phi_CS}
    \phi(r)=
    \begin{cases}
        \phi_{s}[r/r_{s}(1+r/r_{s})^{2}] \equiv \phi_{int}(\approx 0)&(r<r_{c})
        \\
        -\frac{\beta\rho_{s}r_{s}^{2}}{M_{\mathrm{Pl}}}\frac{\mathrm{ln}(1+r/r_{s})}{r/r_{s}}-\frac{C}{r/r_{s}}+\phi_{\infty} \equiv \phi_{out}& (r>r_{c})
    \end{cases}
\end{equation}
where 
\begin{equation}
    C \approx -\frac{\beta\rho_{s}r_{s}^{2}}{M_{\mathrm{Pl}}}\mathrm{ln}(1+r_{c}/r_{s})+\phi_{\infty}r_{c}/r_{s},
\end{equation}
\begin{equation}
    \phi_{\infty}-\frac{\beta\rho_{s}r_{s}^{2}}{M_{\mathrm{Pl}}}(1+r_{c}/r_{s})^{-1}\approx 0.
\end{equation}

\textit{Vainshtein Screening}: In the case of Vainshtein screening the additional degrees of freedom are screened through a non-linear mechanism. In Vainshtein screening the gravitational potentials are modified inside the matter sources. These modifications are screened outside the sources and GR is recovered in low-density environments, as required by tests of gravity. GR is not recovered everywhere within the Vainshtein radius as the screening breaks down there. The Vainshtein radius $r_{V}$ is given by the curvature of the object \cite{Paillas_2019, Brando_2023}.


The modified gravitational potential is now given as \cite{Cardone2020Aug, Haridasu:2021hzq, Crisostomi_2018, Dima_2018, Langlois_2018, Bartolo_2018},
\begin{equation}
\label{eqn:grav_vain}
    \frac{\diff \Phi(r)}{\diff r}=\frac{G_{\mathrm{N}}^{\mathrm{eff}}{M}(r)}{r^{2}}+\Xi_{1}G_{\mathrm{N}}^{\mathrm{eff}}{M}''(r)
\end{equation}
where $G_{\mathrm{N}}^{\mathrm{eff}}$ is the modified Newton's constant defined as $G_{\mathrm{N}}^{\mathrm{eff}}=\tilde{\gamma}_{\mathrm{N}} \times \GN$ and $'$ represents the derivative w.r.t $r$. The effective parameter of the theory $\Xi_1$ is related to the physical parameters as \cite{Langlois_2017,Cardone2020Aug},
\begin{equation}
    \Xi_{1} =- \frac{(\alpha_{1}+\beta_{H})^{2}}{2(\alpha_{H}+2\beta_{1})}
\end{equation}

In \cite{Haridasu:2021hzq}, we set $\gNtilde =1$ when presenting the final results, as this parameter will be completely degenerate with normalization of mass term ${M}(r)$. Hence, the assessment was limited only to the parameter $\Xi_1$ as the physical parameters can only be disentangled when the weak lensing potential is utilized complementary to the hydrostatic potential \cite{Cardone2020Aug, Haridasu:2021hzq}, in the case of Vainshtein screening, the gravitational potential and the weak lensing potential are different. In the current work, although a joint analysis with different mass-estimating techniques is utilized, the potential is the same, unable to break additional degeneracies. However, we anticipate that the improvement in the constraints of mass should help with better constraining the $\Xi_1$ parameter. Although the $\gNtilde$ is unconstrained we leave it as a free parameter when performing the joint importance sampling analysis, to assess the improvement in the constraints. Note that both for the Chameleon and the Vainshtein screening mechanisms we are solely interested in the modifications to the gravitational potential given by \Cref{eqn:grav_vain,eqn:grav_cham} to perform the comparisons with the caustic surface estimated through the phase space of galaxies. 

\section{Data}
\label{sec:data}

In this work, we analyze X-ray and spectroscopic data of two massive galaxy clusters, A2029 and A2142 at redshifts, $z = 0.0784$ and $z = 0.08982$, respectively. 

\textit{Hydrostatic data}: These clusters have been amply studied, establishing their hydrostatic masses and other thermodynamic properties, within the X-COP compilation presented in \cite{Ettori:2018tus, Eckert:2018mlz, Ghirardini:2018byi}, which have been earlier utilized to test modified gravity scenarios \cite{Ettori:2018tus, Haridasu:2021yrw, Haridasu:2021hzq, Gandolfi:2023hwx, Benetti:2023vxy}. Note that these two clusters are relaxed \cite{Ettori:2018tus} and have no non-thermal pressure support at large radii \cite{Eckert:2018mlz}. 

\textit{Caustic data}: Meanwhile, the necessary data required for estimating caustics are sourced from two references, \cite{Sohn_2017} for A2029 and \cite{Liu_2018} for A2142. These datasets, detailed in \Cref{fig:causticprofiles-A2029} and \Cref{fig:causticprofiles-A2142}, consist of galaxy right ascension, declination, and redshift information within the A2029 and A2142 clusters. This data forms the foundation for constructing the velocity profiles of these clusters, as visually represented in \Cref{fig:causticprofiles-A2029} and \Cref{fig:causticprofiles-A2142}. For A2142 we have 2239 galaxies and for A2029 we have 982 galaxies. The membership selection of these galaxies for the formation of the caustics is done using the method detailed in \cite{Gifford13}. {Note that these two clusters also have very high $\sim 95\%$ sample completeness of the member galaxies, which allows for a more accurate estimation of the caustic surface}.

In effect, we choose to work with these two clusters as they have very good hydrostatic and caustic data simultaneously. {The analysis can be extended to clusters that have both the dynamics and kinematics observed, however subject to the precision of the individual measurements (for example, the Coma cluster \cite{Sohn_2017}). We anticipate the possibility of extending the analysis to a larger sample \cite{Logan_2022}, where 44 galaxy clusters were used to estimate the hydrostatic and caustic bias. Improved hydrostatic observations NIKA2 \cite{Ruppin:2017bnt} and possible improvements of the observed kinematics would benefit from the application of the method implemented here.} To calculate the gravitational potentials associated with GR, CS, and VS models, as well as the resulting profile $\mathcal{F}_{\beta}(r)$ from the hydrostatic data, we employ the Markov Chain Monte Carlo (MCMC) chains, as utilized to constrain the models in \cite{Boumechta:2023qhd, Haridasu:2021hzq, Haridasu:2021yrw}. These chains contain all the essential parameters such as $M_{200}^{\mathrm{HS}}$, $R_{200}^{\mathrm{HS}}$, $c_{200}^{\mathrm{HS}}$, and the modified gravity parameters.

\section{Method}
\label{sec:Analysis}

Under the assumption that the cluster is mostly dominated by dark matter, specifically in the range beyond $\sim 30 \, \mathrm{kpc}$ which we consider, we can model the mass density using the NFW profile \cite{Schaller15, Hogan2017, Sartoris2020}, which is given as \cite{Navarro:1995iw},
\begin{equation}
    \label{eqn:NFW_density_profile}
    \rho(r)=\frac{\rho_{s}}{(r/r_s)(1+r/r_s)^{2}}
\end{equation}
where $\rho_{s}$ is the characteristic density, $r_{s}$ is the characteristic radius, where the logarithmic slope $s=d \mathrm{ln}\rho/d \mathrm{ln}r$ takes the isothermal value s=-2. The corresponding mass profile \cite{Cardone2020Aug,Haridasu:2021hzq} in terms of the $\Md$ and concentration $\cd$ is given as,
\begin{equation}
    \label{eqn:NFW_mass_profile}
    M(<r)=\Md \frac{\mathrm{ln}(1+\cd x)-\cd x/(1+\cd x)}{\mathrm{ln}(1+\cd)-\cd/(1+\cd)}
\end{equation}
where $x=r/\Rd$, $\cd=\Rd/r_{s}$ and 
\begin{equation}
    \Md=\Delta \frac{4}{3}\pi\rho_{c}(z)\Rd^{3}.
\end{equation}

We appropriately assume the value of the $\Delta$, to be to the usual value $\Delta = 200$ when making comparisons. we utilize the NFW mass profile to assess the masses in the GR scenario and also in the Vainshtein and Chameleon screening scenarios, as we have earlier done in \cite{Haridasu:2021yrw, Haridasu:2021hzq} and  \cite{Boumechta:2023qhd}, respectively. {The assumption of NFW profile in the current implementation is supported by the fact that it is the best-fitting profile in the GR case \cite{Ettori_2019}. As the caustic technique is essentially independent of such an assumption on the mass profile, it could be an interesting exercise to assess the same for different assumptions of the mass profile, which we leave for a future discussion. In our preliminary assessment, we find that a varied assumption of the mass profile, for example, Burkert \cite{Burkert01}, makes little to no difference in the final constraints obtained on the modified gravity parameters. }

\subsection{Computing $\mathcal{F}_{\beta}(r)$}

\begin{figure}[!ht]
    \centering
    \includegraphics[scale=0.45]{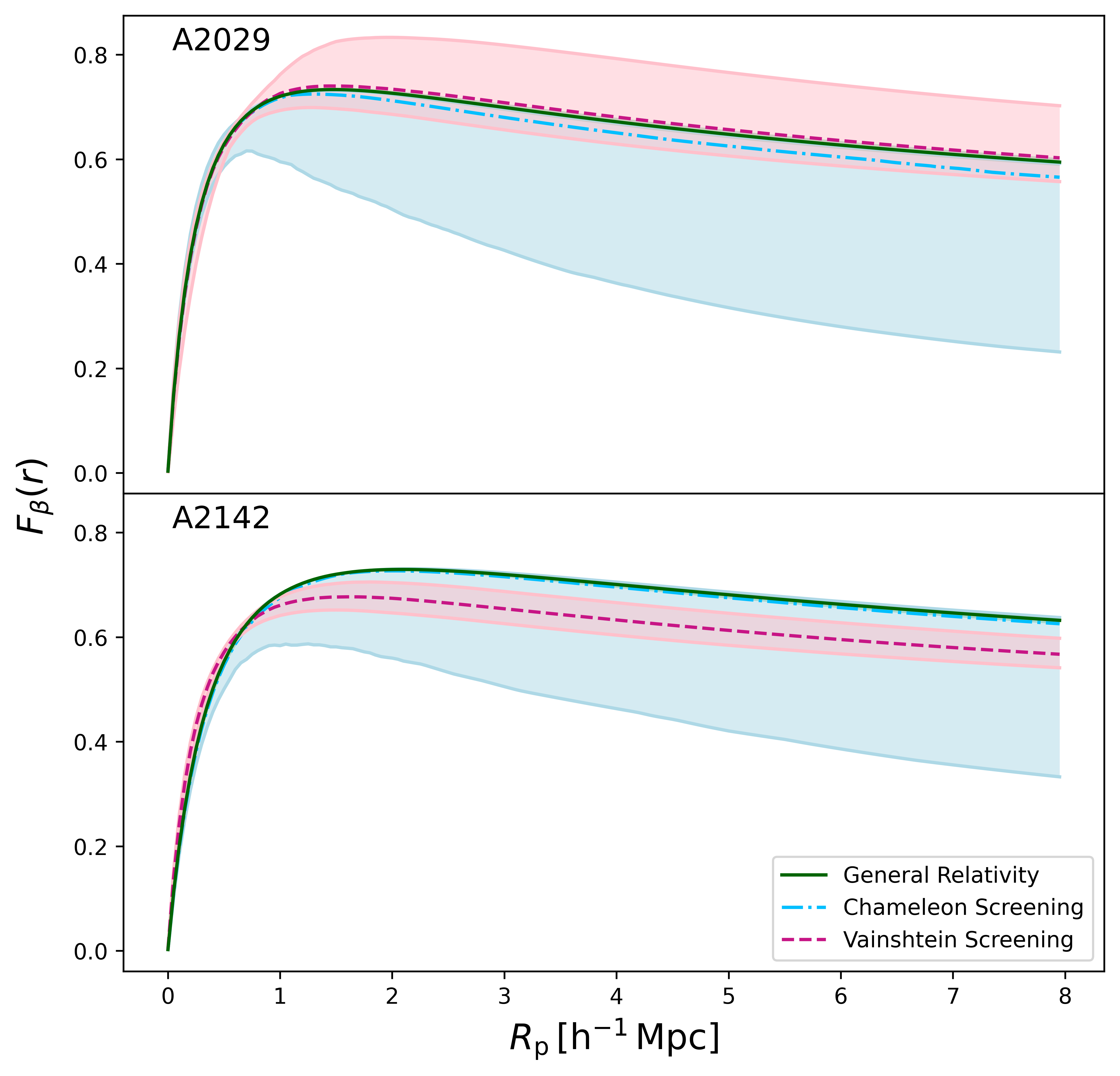}
    \caption{ We show the profiles of $\fbr$ constructed utilizing the constraints on the gravitational potential obtained by fitting the hydrostatic data. In the \textit{top} we show the profiles for the cluster A2029, for GR (green), Chameleon screening (blue), and Vainshtein screening (red). Similarly, in \textit{bottom} panel we show the same for the cluster A2142. {Note that the GR case coincides with the upper limit of the Chameleon scenario, appearing as an overlap of the lines. } }
    \label{fig:fbr}
\end{figure}

In the computation of $\mathcal{F}_{\beta}(r)$, we rely on the expressions detailed in \Cref{eqn:fofr} and \Cref{eqn:gbeta}. Here, $\rho(r)$, as defined in \Cref{eqn:NFW_density_profile}, serves as a crucial component and $\Phi(r)$ represents the gravitational potential. In the context of General Relativity (GR), $\Phi(r)$ corresponds to the familiar Newtonian potential. In the Chameleon Scalar-Tensor (CS) and Vector Scalar-Tensor (VS) models, $\Phi(r)$ assumes the forms outlined in \Cref{eqn:grav_cham} and \Cref{eqn:grav_vain}, respectively.\\
We perform the calculation for $\mathcal{F}_{\beta}(r)$ across 1000 profiles, each corresponding to distinct parameter values drawn from the MCMC\footnote{We utilize the \texttt{emcee} code \cite{Foreman-Mackey13} publicly available at \href{https://github.com/dfm/emcee}{https://github.com/dfm/emcee}.} samples obtained after fitting the hydrostatic data, see \cite{Boumechta:2023qhd} for Chameleon, and \cite{Haridasu:2021hzq} for Vainshtein screening. In the resulting plot depicted in \Cref{fig:fbr}, the solid lines represent the mean values extracted from these profiles, while the shaded region encompasses the $95\%$ C.L. allowed regions of the profiles.

\subsection{Measuring $\mathcal{A}(r)$}

To determine the caustic amplitude, denoted as $\mathcal{A}^{2}(r)$, we employ a methodology based on the velocity dispersion profile derived from observational data. This technique, as described in \cite{Diaferio99}, relies on the velocity distribution of galaxy clusters. We begin by utilizing velocity data obtained from the members of galaxy clusters, as provided in the datasets \cite{Sohn_2017} and \cite{Liu_2018} for A2029 and A2142 respectively. Following the selection of velocities \cite{Gifford13}\footnote{We use part of the code provided in \cite{Gifford13} to produce the caustic surfaces and extract the caustic amplitude for the mass calculation.}, taking into account the full phase-space \cite{Wing_2011} for inclusion in the caustic diagram, we proceed to define the caustic surfaces using the technique detailed in \cite{Diaferio99}. Consider a collection of $N$ galaxies, each characterized by coordinates $\mathbf{x} = (r, v)$, where we conveniently rescale both $r$ and $v$. We adopt an adaptive kernel method to estimate the density distribution of these galaxies within the redshift diagram.


This density is then expressed as
\begin{equation}
    \label{eqn:density_distribution}
    f_{q}(\mathbf{x})=\frac{1}{N}\sum_{i=0}^{N}\frac{1}{h_{i}^{2}}K\left(\frac{\mathbf{x}-\mathbf{x}_{i}}{h_{i}}\right).
\end{equation}
And the kernel function $K(t)$ is defined as
\begin{equation}
    K(t)=
    \begin{cases}
        4\pi^{-1}(1-t^{2})^{3} & t<1
        \\
        1 & \mathrm{otherwise} .
    \end{cases}
\end{equation}

The local smoothing length, denoted as $h_{i}$, depends on both the local density and an optical smoothing length $h_{\mathrm{opt}}$, which is given by:
\begin{equation}
  h_{\mathrm{opt}}=\frac{3.12}{N^{1/6}} \left( \frac{\sigma_{r}^{2}+\sigma_{v}^{2}}{2}\right)^{1/2} .
\end{equation}
Here, $\sigma_{r}$ and $\sigma_{v}$ represent the marginal standard deviations of the galaxy coordinates. Additionally, a local smoothing factor $\lambda_{i}$ is introduced as $\lambda_{i} = [\gamma/f_{1}(\mathbf{x}{i})]^{1/2}$, where $f_{1}$ is calculated from the density distribution. Notably, we set $h_{c} = \lambda_{i} = 1$ for any $i$, and $\log \gamma$ is computed as $\sum_{i} \log[f_{1}(\mathbf{x}_{i})]/N$.

The amplitude $\mathcal{A}^2(r)$ at a fixed radial distance $r$ is then determined by solving the equation $f_{q}(r, v) = \kappa$. More specifically, the first upper and lower solutions, denoted as $v_{\mathrm{u}}$ and $v_{\mathrm{d}}$, are identified away from the maximum of $f_{q}(r, v)$, closest to $v = 0$. The amplitude $\mathcal{A}(r)$ is computed as $\min\{|v_{\mathrm{u}}|, |v_{\mathrm{d}}|\}$. It's worth noting that the prescription $\mathcal{A}(r) = \min{|v_{u}|, |v_{d}|}$ is equivalent to $\mathcal{A}(r) = (v_{\mathrm{u}} - v_{\mathrm{d}})/2$ in the case of an isolated spherically symmetric system.

While we obtain $f_{q}(r, v)$ uniquely except for the choice of $q$ \footnote{The value of $q$ could be assumed between 5-10, and we validate that it does not affect the estimation of the caustic profile, within the estimated error. In the current work, we utilize $q = 10$, finding no major difference with the assumed value. }, it's important to highlight that there exists an infinite range of thresholds $\kappa$ that can be employed to determine $\mathcal{A}(r)$. In practice, it's reasonable to assume that, particularly in the central region, the cluster has reached a state of virial stability. Thus, within this central region, the equation $\langle v^{2}_{\mathrm{esc}}\rangle_{R} = \langle v^{2}\rangle_{R}$ holds. Here, the angular brackets indicate an average computed over the entire sphere of radius $R$, considering velocities in three dimensions. In our specific dataset, for example, from \cite{Sohn_2017}, we only have access to one-dimensional velocity information. Therefore, the methodology also assumes that if the velocity field exhibits approximate isotropy in the central region, our expression remains valid. Consequently, we have $\langle v^{2}\rangle_{\kappa,R} = \int_{0}^{R}\mathcal{A}^{2}(r)\Phi(r)dr / \int_{0}^{R} \Phi(r) dr$, where $\Phi(r)$ represents the integral of $f_{q}(r, v)$. It's important to emphasize that $\langle v^{2}_{\mathrm{esc}}\rangle_{\kappa, R}$ is the only quantity that depends on the chosen threshold $\kappa$.

To select an appropriate $\kappa$, we minimize the function defined as:
\begin{equation}
    \label{eqn:kappa_eqn}
    S(\kappa,R)=|\langle v_{\mathrm{esc}}^{2}\rangle_{\kappa,R}-4\langle v^{2}\rangle_{R}|^{2}.
\end{equation}
Here, we define the radial distance $R$ as the mean projected distance of cluster members from the cluster center, and $\langle v^{2}\rangle_{R}$ represents the one-dimensional velocity dispersion of the cluster members. This comprehensive methodology enables us to robustly estimate the caustic amplitude and, by extension, gain insights into the cluster's dynamical properties and mass distribution.

The uncertainty in the measured value of $\mathcal{A}(r)$ depends on the number of galaxies contributing to the determination of $\mathcal{A}(r)$ \cite{Serra_2010}. Therefore, we define the relative error,
\begin{equation}
    \label{eqn:error_A(r)}
    \delta\mathcal{A}(r)/\mathcal{A}(r)=\kappa/\mathrm{max}\{f_{q}(r,v)\}
\end{equation}
where the maximum value of $f_{q}(r, v)$ is along the $v$-axis at fixed $r$. The resulting error on the cumulative mass profile is
\begin{equation}
    \label{eqn:error_Mass}
    \delta M_{i}=\sum_{j=1}^{i}|2m_{j}\delta\mathcal{A}(r_{j})/\mathcal{A}(r_{j})|
\end{equation}
where $m_{j}$ is the mass of the shell $\{r_{j-1}, r_{j}\}$ and $i$ is the index of the most external shell.

\subsection{Comparing $\mathcal{A}^{2}(r)$}

As a preliminary assessment, we utilize \Cref{eqn:phi_caustic}, to perform a comparison between the gravitational potential on the L.H.S estimated using the hydrostatic equilibrium and the caustic surface on R.H.S, using the phase space data of the galaxies. 
The fundamental definition of $\mathcal{A}^{2}(r)$ is articulated in \Cref{eqn:phi_caustic}. In our investigation, we undertake a comparative study involving caustic profiles derived from empirical observations and theoretical projections of $\mathcal{A}^{2}(r)$. The latter is determined through both \Cref{eqn:phi_caustic} and the inversion of \Cref{eqn:Mass_Profile}. While \Cref{eqn:phi_caustic} yields a prediction for $\mathcal{A}^{2}(r)$ independent of $\mathcal{F}_{\beta}(r)$, \Cref{eqn:Mass_Profile} provides predictions contingent upon $\mathcal{F}_{\beta}(r)$. We meticulously scrutinize how $\mathcal{A}^{2}(r)$ behaves under constant and varying $\mathcal{F}_{\beta}(r)$ scenarios. Additionally, these equations allow us to quantitatively assess the fluctuations in $\mathcal{A}^{2}(r)$ within the context of modified gravitational potentials while assuming an NFW mass profile, as defined in \Cref{eqn:NFW_mass_profile}. This approach empowers us to estimate the requisite mass adjustments to align a modified gravitational potential with the observed $\mathcal{A}^{2}(r)$. In essence, this methodology provides an indirect route for evaluating the impact of modified gravity on mass bias, an essential aspect of our study, serving as both a justification and validation of our findings. A direct method for obtaining caustic mass involves the straightforward application of \Cref{eqn:Mass_Profile}. This approach constitutes the core of our caustic mass determination. 

The computation of caustic profiles and the assessment of caustic amplitude primarily rely on the velocities and velocity dispersion of galaxy clusters, as outlined by \cite{Gifford13}. Detailed insights into the methodology for obtaining caustic amplitude, denoted as $\mathcal{A}^{2}(r)$, can be found in \cite{Diaferio99} and \cite{Gifford13}. In the existing literature, $\mathcal{F}_{\beta}(r)$ is typically approximated as either 0.5 \cite{Diaferio99} or 0.65 \cite{Gifford13}. In contrast, we implement a more comprehensive approach by utilizing the complete expression for $\mathcal{F}_{\beta}(r)$ involving \Cref{eqn:fofr} and \Cref{eqn:gbeta} apriori derived independently from the hydrostatic data, which in turn is the important aspect of our joint analysis. 

\begin{figure*}
    \centering
    \includegraphics[scale=0.33]{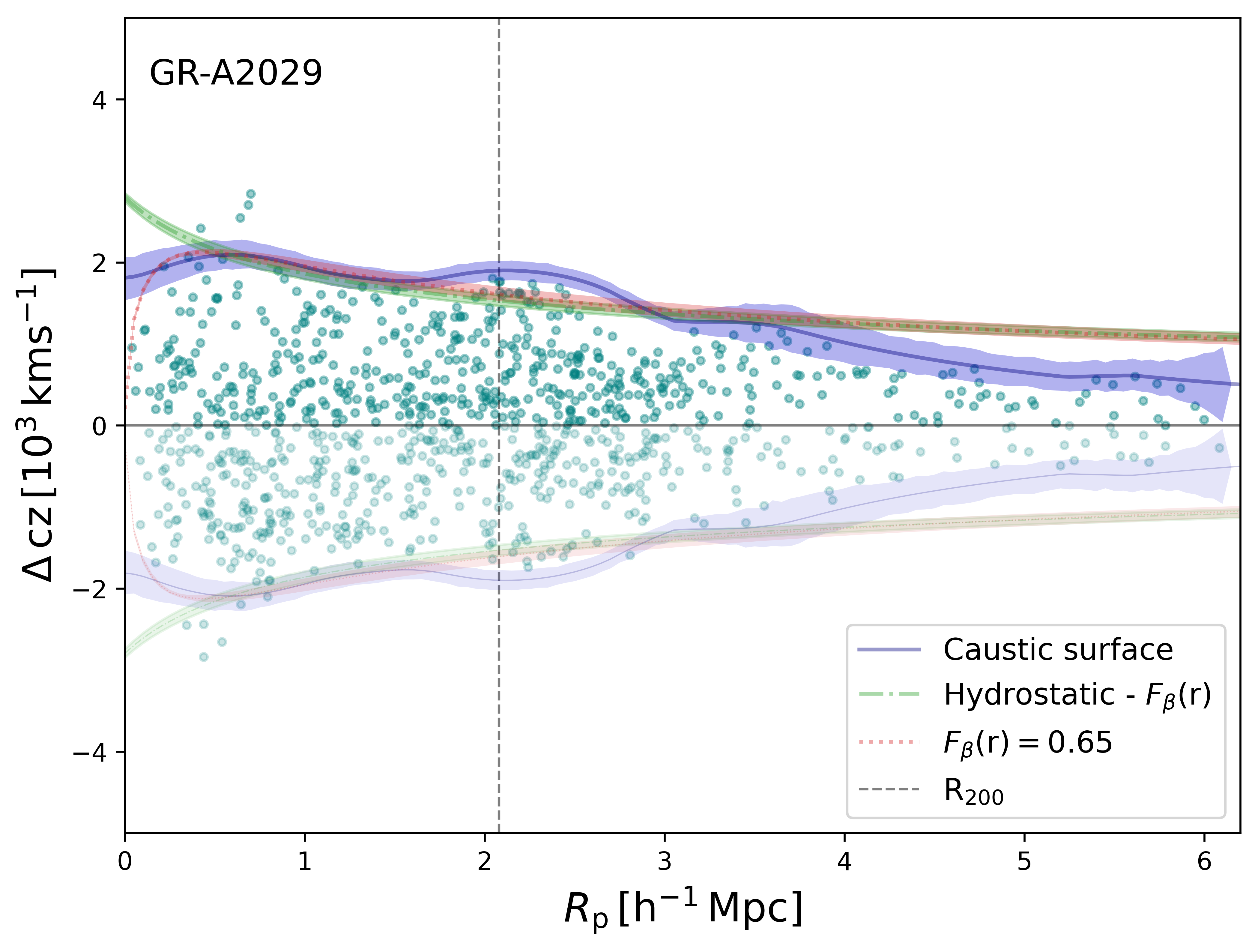}
    \includegraphics[scale=0.33]{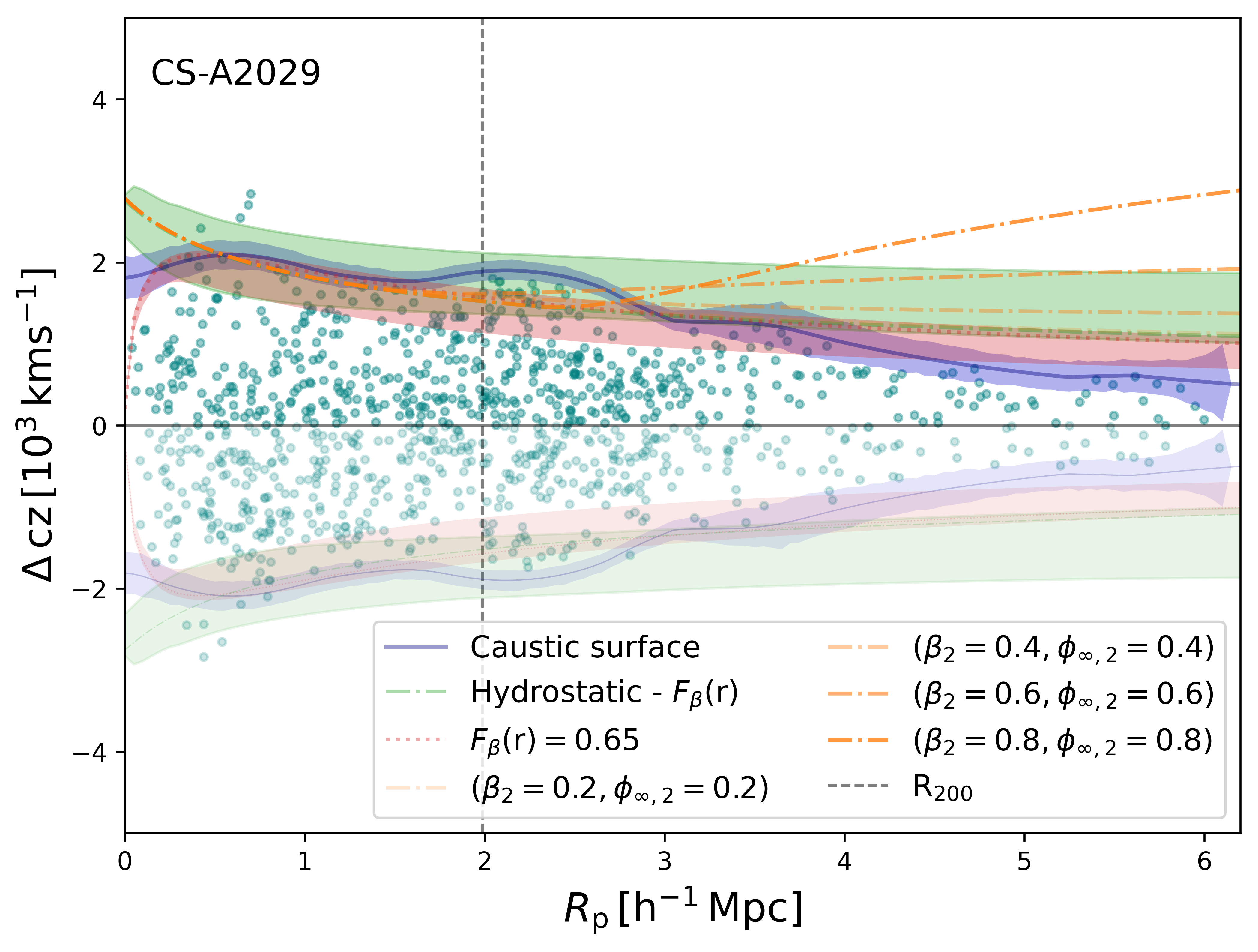}
    \includegraphics[scale=0.33]{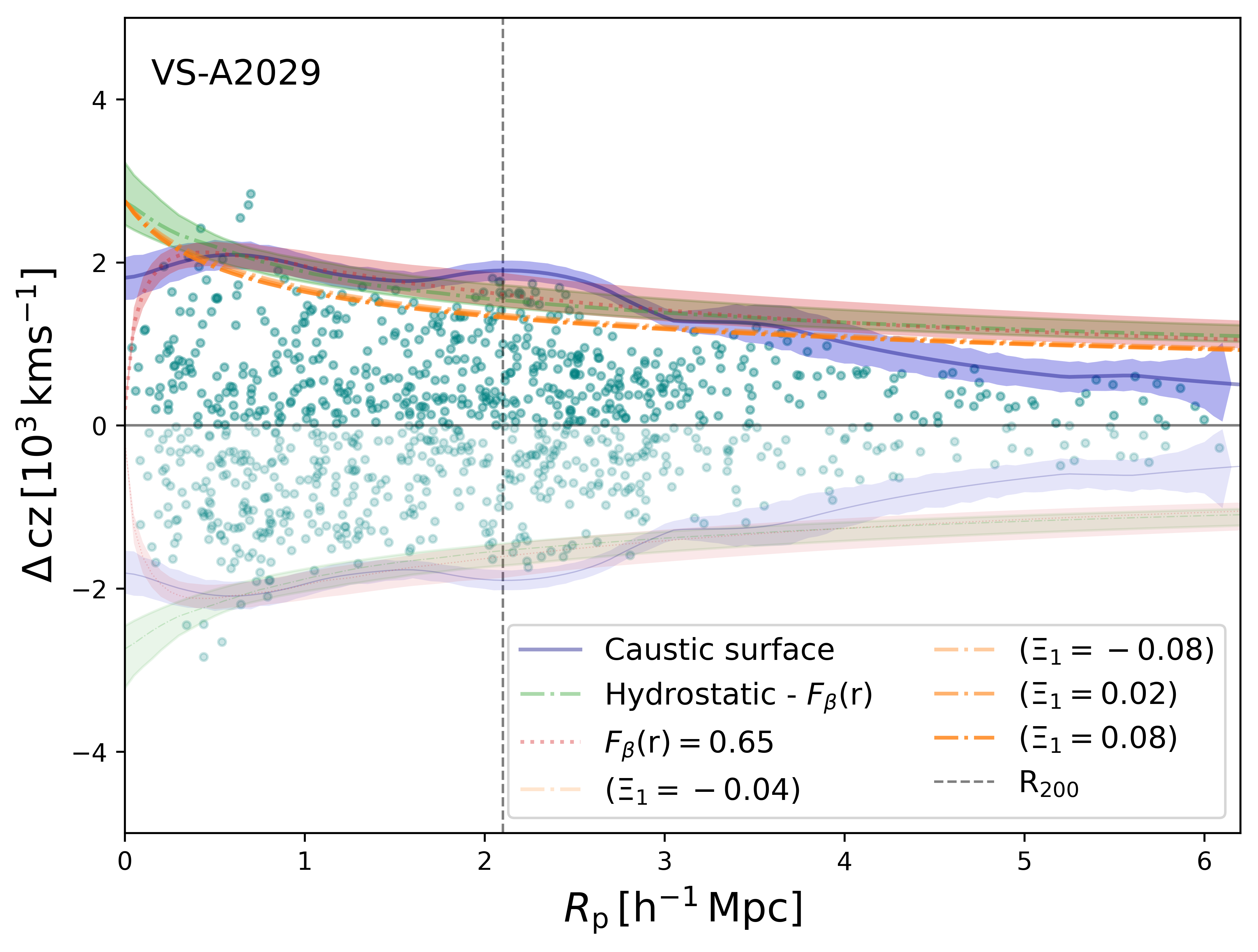}
    \caption{Caustic Profiles for the cluster A2029. We show the caustic profile and the corresponding uncertainty in shaded blue regions, alongside the phase space distribution of the galaxies. \textit{Left}: We compare the caustic profiles estimated using the hydrostatic data in the case of GR, shown in green. {The caustic profile estimated using the hydrostatic data and a constant $\fbr = 0.65$ is shown in red }\textit{Center}: Same as the \textit{Left} panel comparing against the caustic surface estimated using the hydrostatic data in the case of chameleon screening. \textit{Right}: Same as the other two panels, comparing the caustic surface in the case of Vainshtein screening. {For the Chameleon and Vainstein screening cases, we show a few expected caustic surfaces for various values of assumed values of the modified gravity parameters.} }
    \label{fig:causticprofiles-A2029}
\end{figure*}

\begin{figure*}
    \centering
    \includegraphics[scale=0.33]{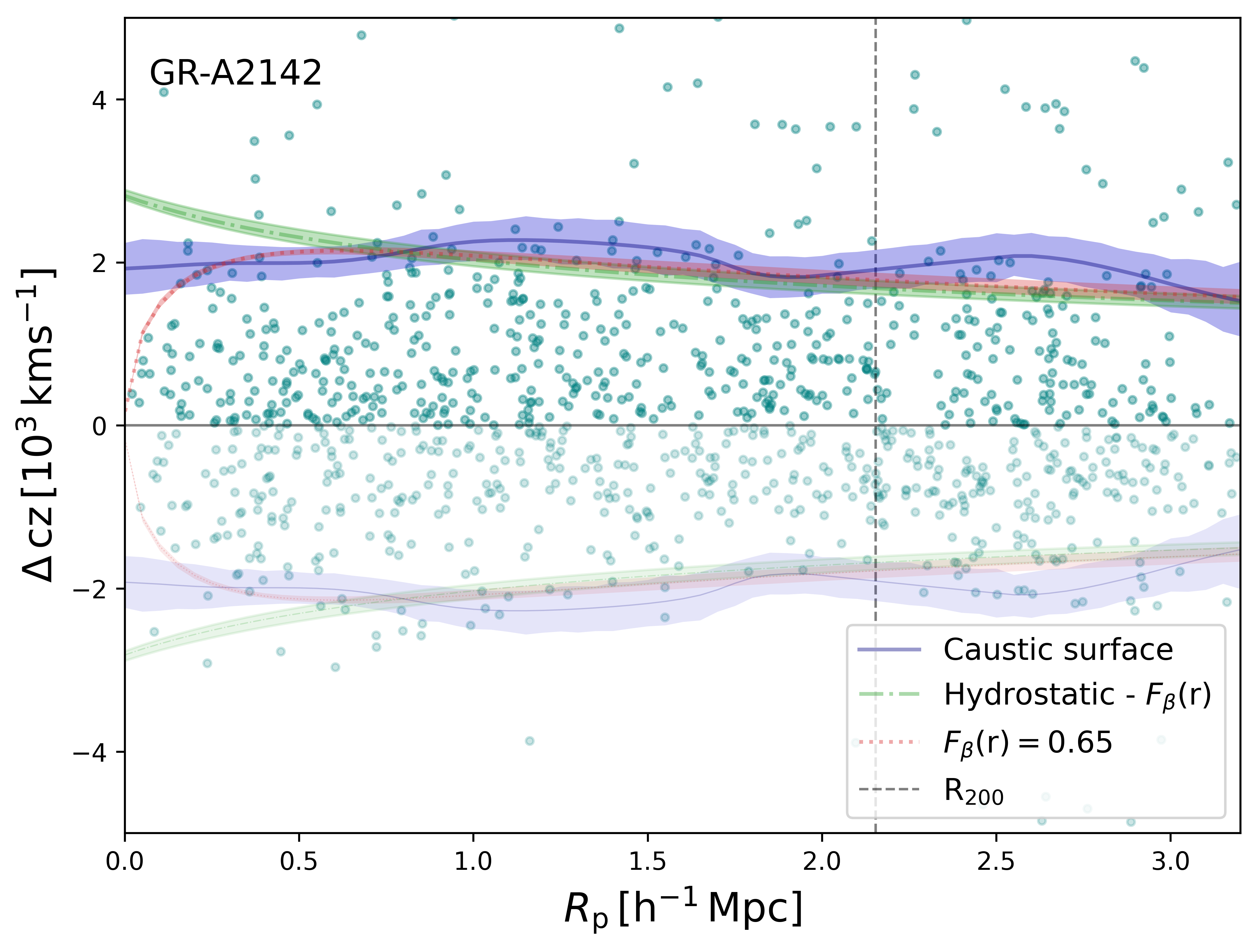}
    \includegraphics[scale=0.33]{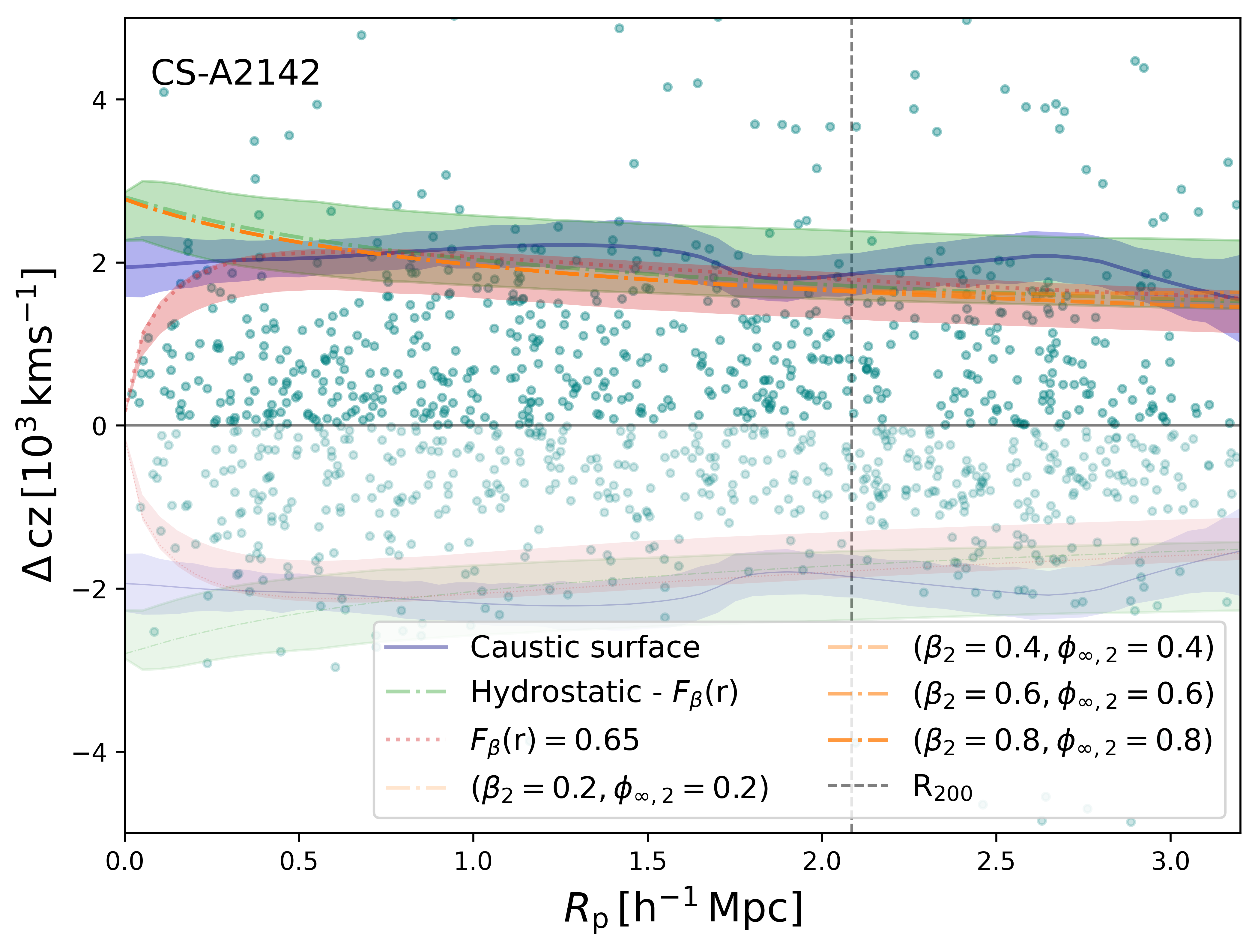}
    \includegraphics[scale=0.33]{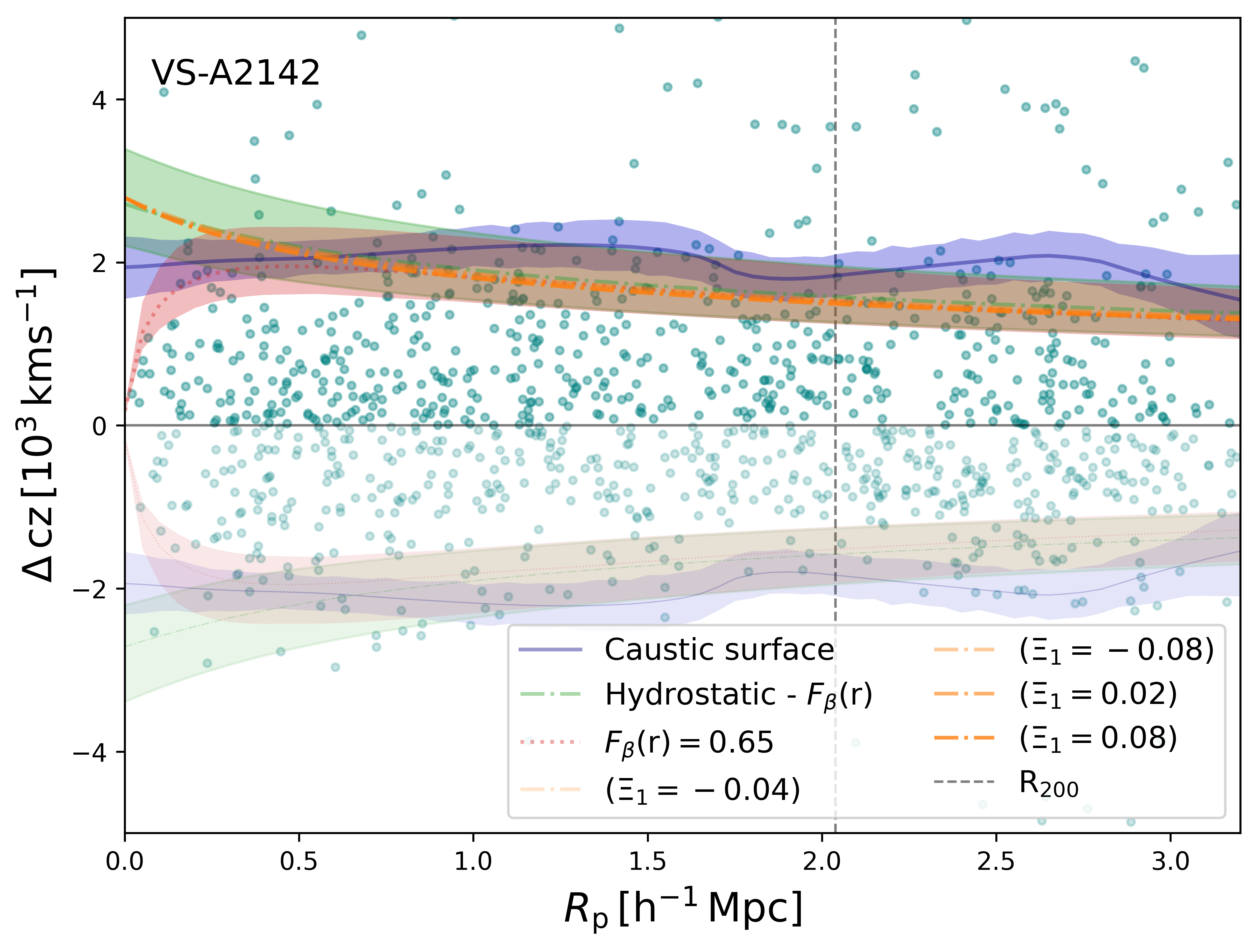}
    \caption{Same as \Cref{fig:causticprofiles-A2029}, but for the cluster A2142. Note the change in the radial range with respect to cluster A2029. }
    \label{fig:causticprofiles-A2142}
\end{figure*}

In GR scenarios, the gravitational potential generated by the cluster, $\Phi(r)$, is commonly known to be \cite{Diaferio99},
\begin{equation}
    \Phi(r)=-\frac{G{M}(<r)}{r}-4\pi G\int^{\infty}_{r}\rho(x)x \diff x
\end{equation}
and $\rho(r)$ is the NFW density profile given in \Cref{eqn:NFW_density_profile}. It is straightforward to evaluate $\mathcal{F}(r)$ in this scenario:
\begin{equation}
    \mathcal{F}_{\mathrm{NFW}}(r)=\frac{r^{2}}{2(r+r_{s})^{2}}\frac{1}{\mathrm{ln}(1+r/r_{s})}
\end{equation}
where $r_{s}$ is the scale radius \cite{Diaferio09}. Furthermore, we extend our analysis to modified gravity scenarios by evaluating $\mathcal{F}_{\beta}(r)$ using our modified potentials, \Cref{eqn:grav_cham} and \Cref{eqn:grav_vain}. This multifaceted approach equips us to comprehensively explore the intricate interplay between mass predictions, gravitational potentials, and modified gravity scenarios.
In \Cref{fig:causticprofiles-A2029} and \Cref{fig:causticprofiles-A2142} we present a comparative analysis of the caustic profiles derived from the two methodologies.
{In each plot, the blue line labels the caustic profile obtained with the caustic technique, while the green line, labeled as Hydrostatic-GR/CS/VS, shows the caustic profile independently predicted by the hydrostatic equilibrium analysis for the three models, respectively.}

Remarkably, the caustic profiles obtained through these two methods exhibit a commendable level of agreement. {The caustic profiles estimated using a constant $\fbr = 0.65$ also agree reasonably well except for the innermost regions. Similarly, an $\mathcal{F}_{\beta}(r) = 0.5$ underestimates the caustic profile, and consequently, we can anticipate that the mass estimated assuming $\mathcal{F}_{\beta}(r)=0.5$ would also be an underestimation. In the current analysis, alongside our method to utilize the hydrostatic data we also assume $\fbr = 0.5 $ or $\fbr = 0.65$ to illustrate the differences and further discuss the implications, especially in the modified gravity scenarios. While an analysis to anticipate a $\fbr$ in the modified gravity scenarios is not present, a recent analysis \cite{Pizzardo:2023idp} in the case of GR find $\fbr = 0.4\pm 0.08$ using simulations \cite{Springel:2017tpz}, which is in agreement with the $0.5$ value. }{To further illustrate potential deviations in the caustic profiles and - by extension - the mass profiles induced in the non-GR case, in the central and left plot of each figure of \Cref{fig:causticprofiles-A2029,fig:causticprofiles-A2142} we additionally show caustic profiles for a few specific values of the modified gravity parameters in both Vainshtein and Chameleon screening.}

\section{Results and Discussion}
\label{sec:results}

We begin by presenting our results for the construction of the caustic profiles for the two clusters considered in our analysis. In \Cref{fig:causticprofiles-A2029,fig:causticprofiles-A2142}, we show the caustic profiles of the clusters A2029 and A2142, respectively. Alongside the reconstructed caustic surface (blue) we also show the caustic surface as anticipated by the hydrostatic technique assuming an NFW mass profile (green), which essentially is a first step validation to further proceed with the joint analysis. In general, we notice a very good agreement between the two estimates. In the \textit{Center} and the \textit{Right} panels we show the same comparison against the Chameleon and Vainshtein screening, respectively. In the case of Chameleon screening, especially for cluster A2029 one can notice a sharp deviation of the caustic surface predicted by hydrostatic equilibrium at the screening radius. It is worth pointing out that this radial range is fairly outside the range covered by hydrostatic data and cannot be constrained, however when compared with the caustic surface one can easily rule out certain parameter space that provides such a large deviation. Essentially, this demonstrates the added advantage of the joint analysis including the caustic technique that can probe the outskirts of the galaxy cluster. In the case of the Vainshtein screening, however, the modifications are not limited to the radial range larger than the screening radius, as the fifth force penetrates the inner regions of the galaxy clusters. This can be easily understood by contrasting the modifications of the gravitational potential in these two modified gravity scenarios (see \Cref{eqn:grav_vain,eqn:grav_cham}). We begin by contrasting our caustic mass estimates to those present in literature to validate our procedure assuming a constant $\fbr$.

\subsection{Joint constraints on Mass}
\label{sec:GR_mass}
\begin{figure*}
    \centering
    \includegraphics[width=0.49\textwidth]{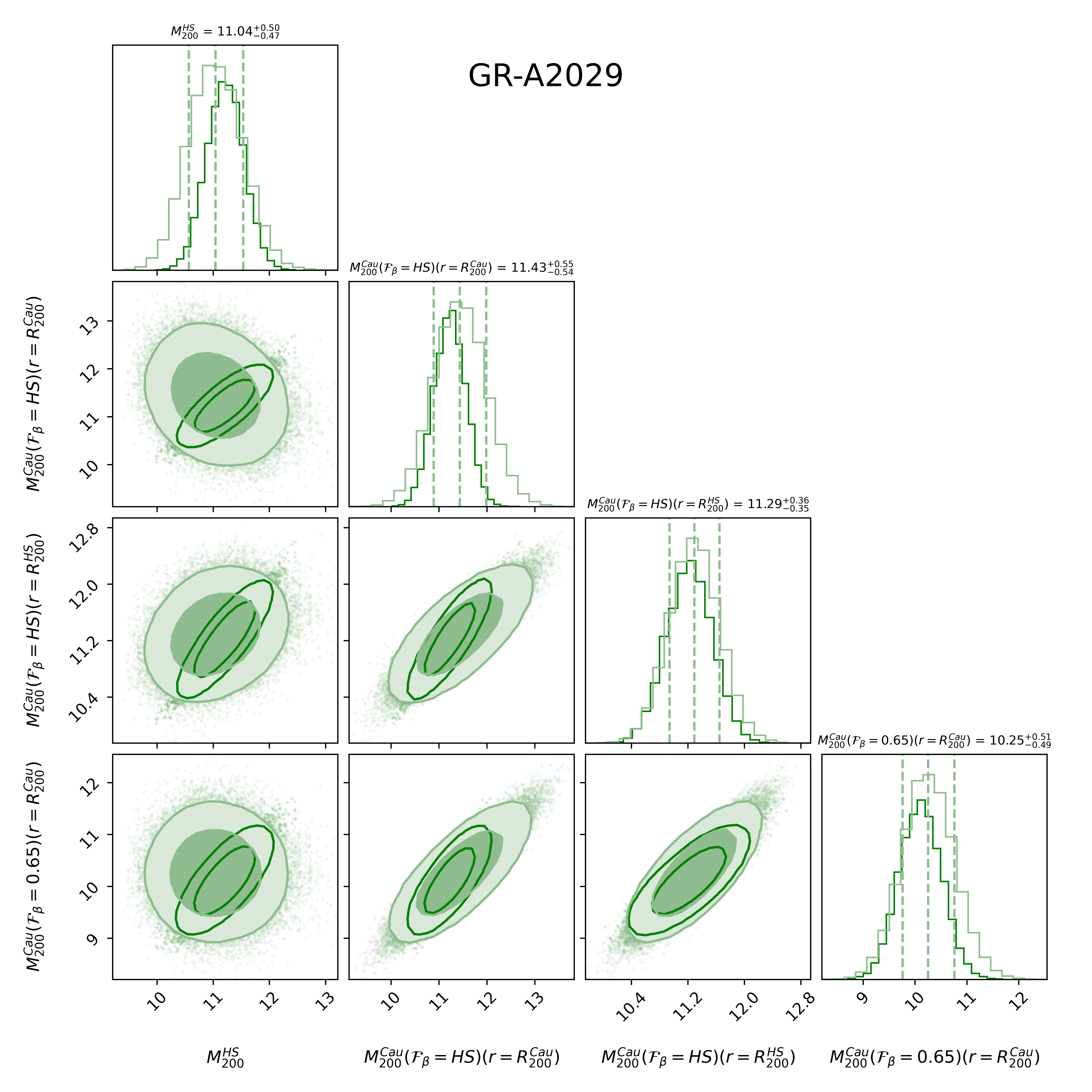}
    \includegraphics[width=0.49\textwidth]{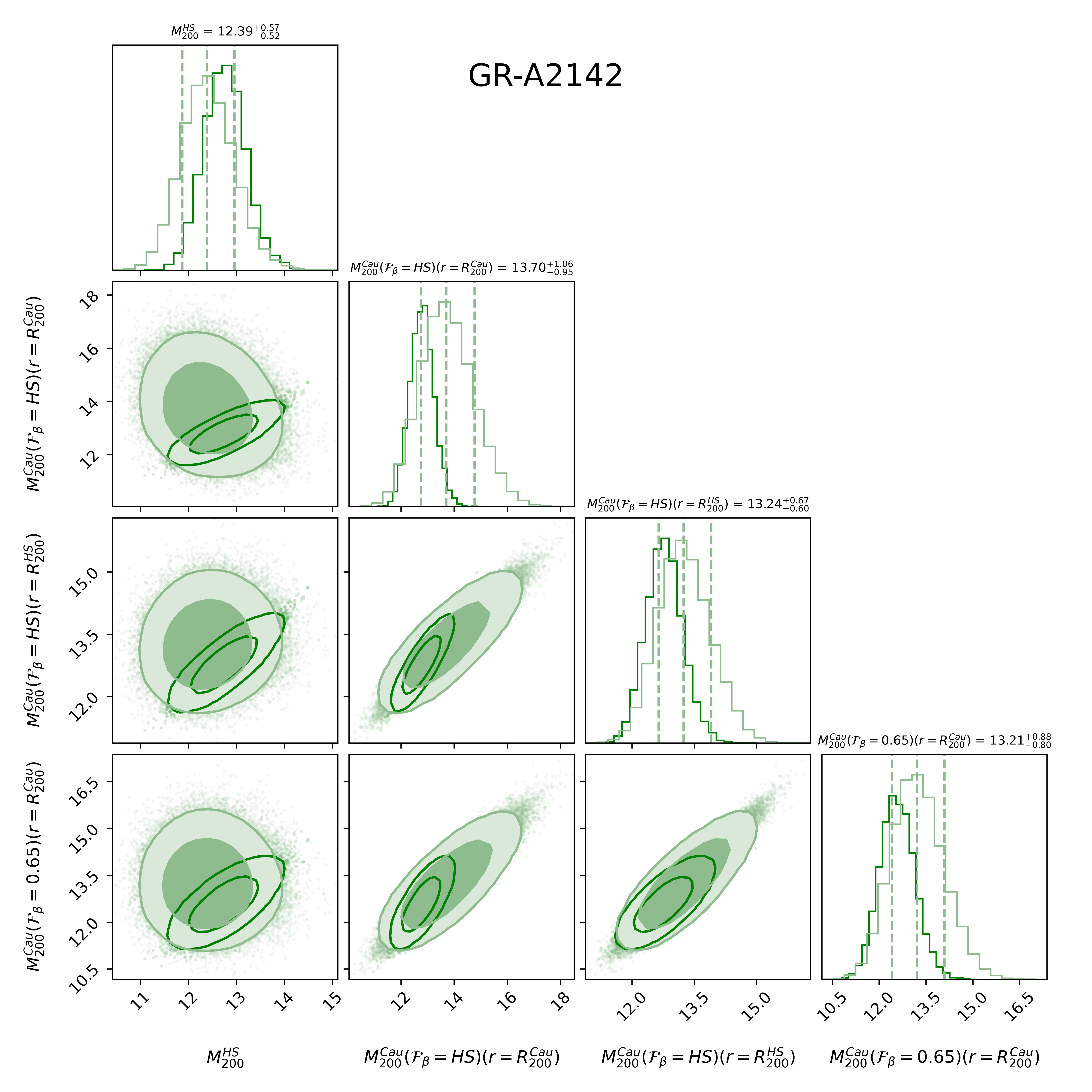}
    \caption{\textit{Left}: We show the $68\%$ and $95\%$, confidence levels of the $M_{200}$ values, estimated using the hydrostatic data, assuming the constant $\fbr = 0.65$ cases. The unfilled contours represent the final importance-sampled posteriors corresponding to the penultimate column of \Cref{tab:GR_Masses}.} 
    \label{fig:Contour_GR}
\end{figure*}

Firstly, we assess the constraint on mass in the standard GR case taking into account the appropriate $\fbr$, well-constrained using the hydrostatic data. In \Cref{fig:Contour_GR}, we show the {marginalized} posteriors for the hydrostatic mass compared against the caustic mass {when assuming a constant} $\fbr$, which are also reported in \Cref{tab:GR_Masses}. As it can be seen from the contours, the caustic mass evaluated at $\Rcau$\footnote{The $\Rcau$ is simply evaluated as the radius within which the average density is 200 times the critical density, after the mass profile is reconstructed.} is mildly anti-correlated with the $\Mhs$, while the mass estimated at $\Rhs$ from the caustic mass profile shows a positive correlation, which is an expected behavior. Once we perform the importance sampling all the mass estimates are positively correlated and heavily affected by the $\fbr$ derived using the hydrostatic data. 

We find that the mass bias earlier quoted in the comparison of the hydrostatic mass and the caustic mass is completely alleviated in our joint analysis. For instance, \cite{Ettori_2019} quote a $6.5\sigma$ and $3.9\sigma$ discrepancy between the two masses, for the clusters A2029 and A2142, respectively. These clusters are among the high significance of deviation analyses within the X-COP compilation. However, this should not come as a surprise as it is well known that the global assumption of $\fbr = 0.5 $ \cite{Diaferio_2005} or $\fbr = 0.65$ \cite{Gifford13}, can lead to such a bias. {As can be seen in \Cref{tab:GR_Masses},  yielding lower mass values, the assumption of $\fbr =0.5$ leads to a larger bias w.r.t. the $\fbr  = 0.65$ case.} When taking the appropriate hydrostatic $\fbr$ we find the mass bias to be no more than $\sim 0.03 - 0.1 $, at a significance of $\lesssim 1\sigma$ for both the clusters (see the last column of \Cref{tab:GR_Masses}). For the sake of comparison, we also show the mass reconstructed using constant $\fbr$ and our method in \cref{fig:mass_profiles}. {Incidentally, the assumption of $\fbr = 0.65$, yields a very good agreement between the independent hydrostatic and caustic mass estimates. In the case of A2142 the assumption of $\fbr = 0.65$ is in fact in slightly better agreement than our analysis using the hydrostatic data. }

\begin{figure}
    \centering
    \includegraphics[width=0.48\textwidth]{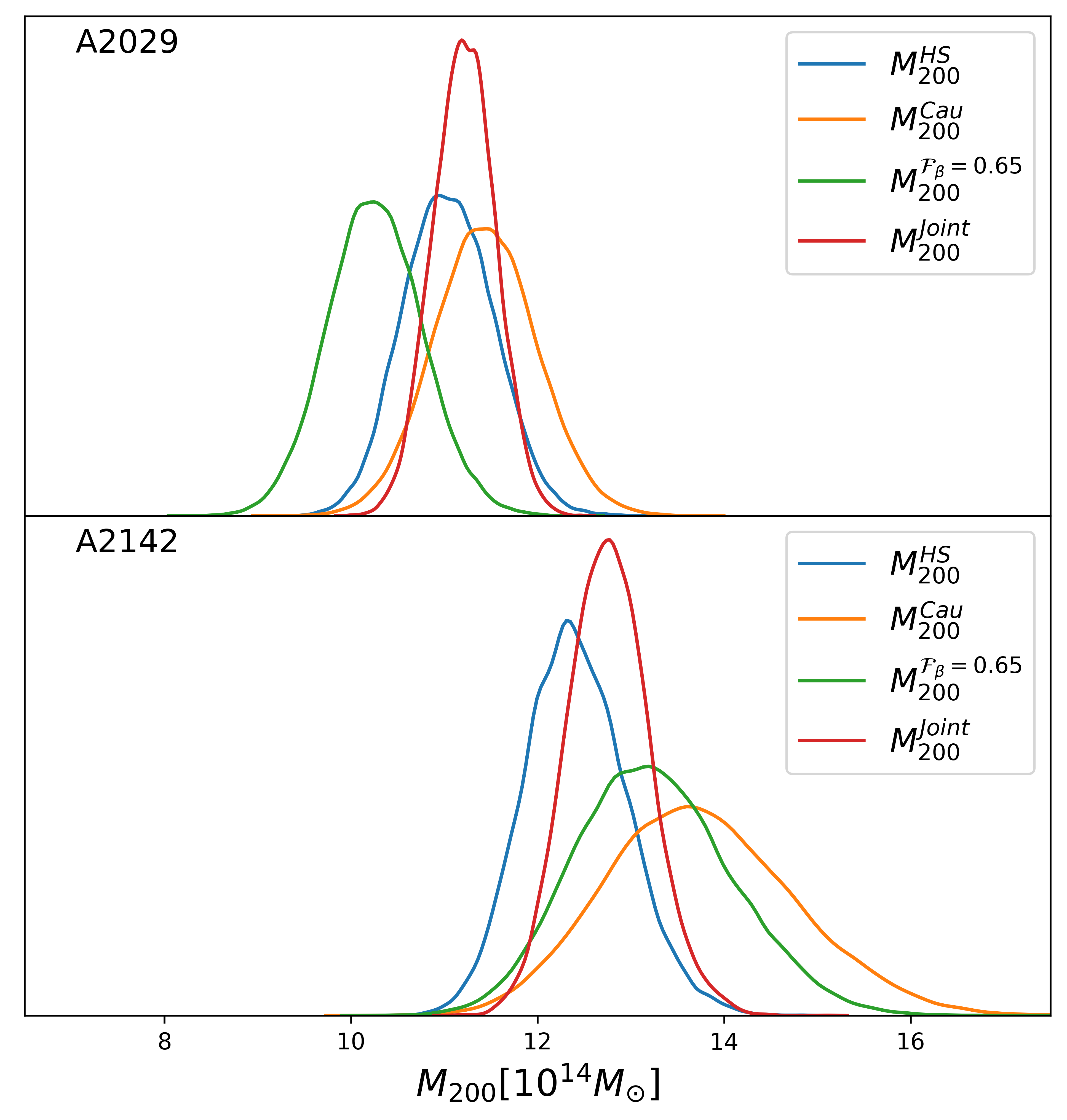}
    \caption{Comparison of the posteriors on the mass estimates from different techniques and the joint analysis, for the clusters A2029 (\textit{top}) and A2142 (\textit{bottom}) panels. }
    \label{fig:mass_comparison}
\end{figure}

On the other hand, we also find that the caustic mass is mildly larger than the hydrostatic masses.  For ease of comparison, we present the posteriors of the masses within the GR case in \Cref{fig:mass_comparison}. Given the very good agreement between the hydrostatic mass and the caustic mass obtained within our formalism, we now perform an importance sampling analysis to obtain the joint constraint on the mass of the cluster. The corresponding values of the constraints on mass are reported in the penultimate column of \Cref{tab:GR_Masses} and shown as red posteriors in \Cref{fig:mass_comparison}. Our joint analysis formalism allows us to obtain tighter constraints on the mass of the clusters. To the best of our knowledge, this is the first combined analysis between these two mass estimation techniques, forward modeling of the hydrostatic data and caustic technique, to obtain a joint constraint that now allows us to extend the formalism to alternate theories of gravity, which indeed is the primary goal of the current investigation.

{\renewcommand{\arraystretch}{1.8}
\setlength{\tabcolsep}{8pt}

\begin{table*}[!ht]
    \caption{Constraints on the mass in the case of GR, shown as $68\%$ C.L. limits. In the second and third columns, we show the constraints obtained using the hydrostatic data only. In columns five and six we quote the caustic mass estimates and in column seven we show the joint mass estimate. In the last column, we show the mass bias between the hydrostatic and caustic techniques. Column 4 shows the corresponding $\fbr$ assumption used in the analysis to estimate the caustic mass. }
    \label{tab:GR_Masses}
    \begin{tabular}{cccccccc}
        \hline
        Cluster & \(\Rhs\)  & \(\Mhs \) & \(\fbr\) & \(\Rcau\) & \(\Mcau \) & \(\Mjoint \) & {\(\Mhs\)}/{\(\Mcau\)} \\
        & [Mpc] & \( [\Munit]\) & & [Mpc] & \( [\Munit]\) & \( [\Munit]\) &  \\
        \hline
        \hline
        \multirow{3}{*}{A2029} &  \multirow{3}{*}{$2.08_{-0.03}^{+0.03}$} &  \multirow{3}{*}{$11.04_{-0.47}^{+0.50}$} & HS & $2.11_{-0.03}^{+0.03}$ & $11.43_{-0.54}^{+0.55}$ & $11.21_{-0.34}^{+0.33}$ & $0.97_{-0.07}^{+0.07}$ \\
        & & & 0.65 & 
        $2.02_{-0.03}^{+0.03}$ &
        $10.25_{-0.49}^{+0.51}$  & $10.79_{-0.38}^{+0.39}$ & $1.08_{-0.07}^{+0.07}$ \\
        & & & 0.50 & $1.86_{-0.03}^{+0.03}$ &
        $7.89_{-0.38}^{+0.39}$  &
        $9.76_{-0.11}^{+0.05}$ & $1.40_{-0.09}^{+0.09}$ \\
        \multirow{3}{*}{A2142} & \multirow{3}{*}{$2.15_{-0.03}^{+0.03}$} & \multirow{3}{*}{$12.39_{-0.52}^{+0.57}$} & HS & $2.23_{-0.03}^{+0.06}$ & $13.70_{-0.95}^{+1.06}$ & $12.75_{-0.43}^{+0.43}$ & $0.90_{-0.08}^{+0.09}$ \\
        & & & 0.65 & 
        $2.20_{-0.04}^{+0.04}$ &$13.20_{-0.79}^{+0.84}$  
        & $12.63_{-0.44}^{+0.47}$& $0.94_{-0.07}^{+0.07}$ \\
        & & & 0.50 & $2.02_{-0.04}^{+0.04}$ &
        $10.16_{-0.61}^{+0.65}$ &
        $11.62_{-0.40}^{+0.41}$ &
        $1.22_{-0.09}^{+0.10}$ \\
        \hline
    \end{tabular}
\end{table*}}

\subsection{Constraints on Modified gravity }
\label{sec:MG_mass}

\begin{figure*}
    \centering
    \includegraphics[width=0.49\textwidth]{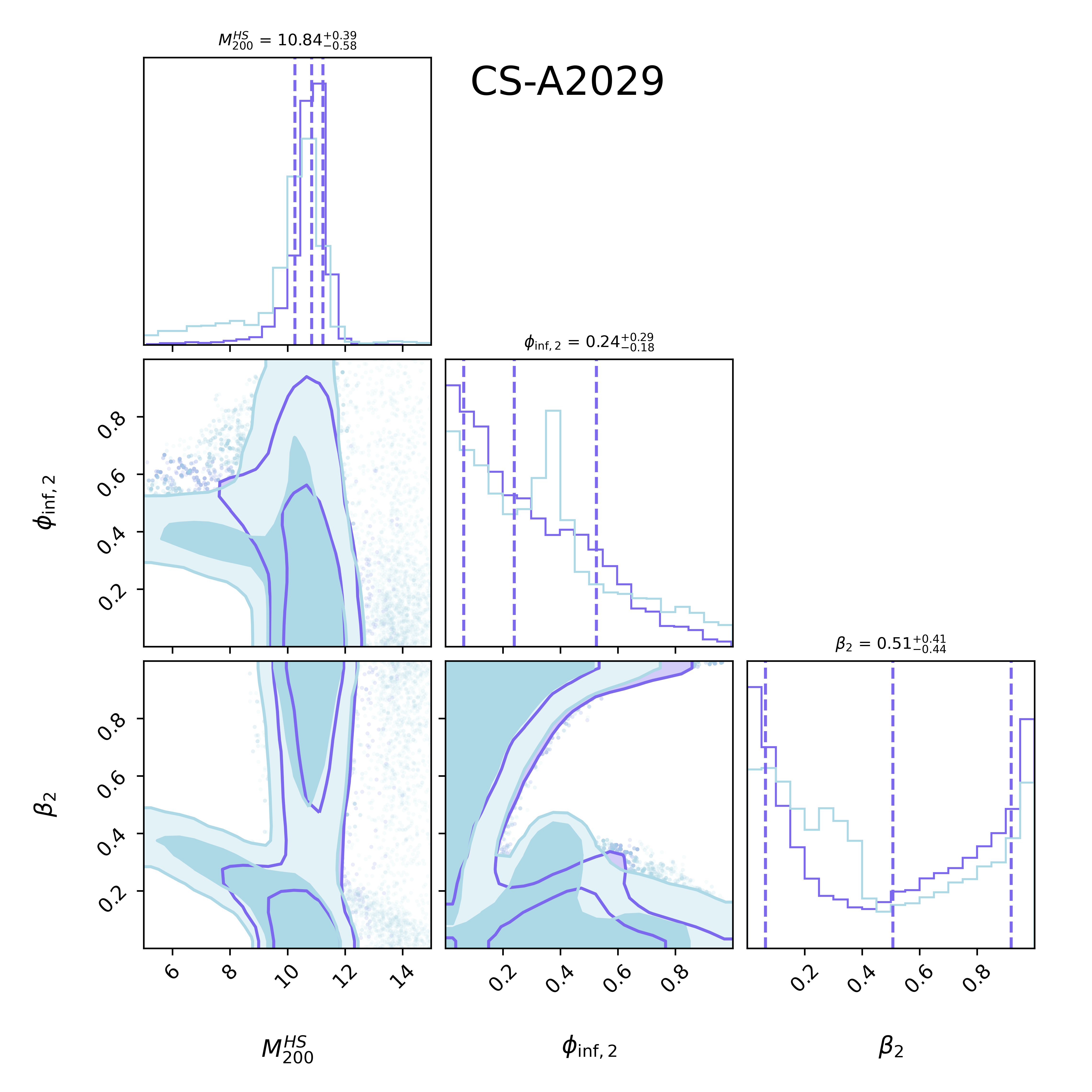}
    \includegraphics[width=0.49\textwidth]{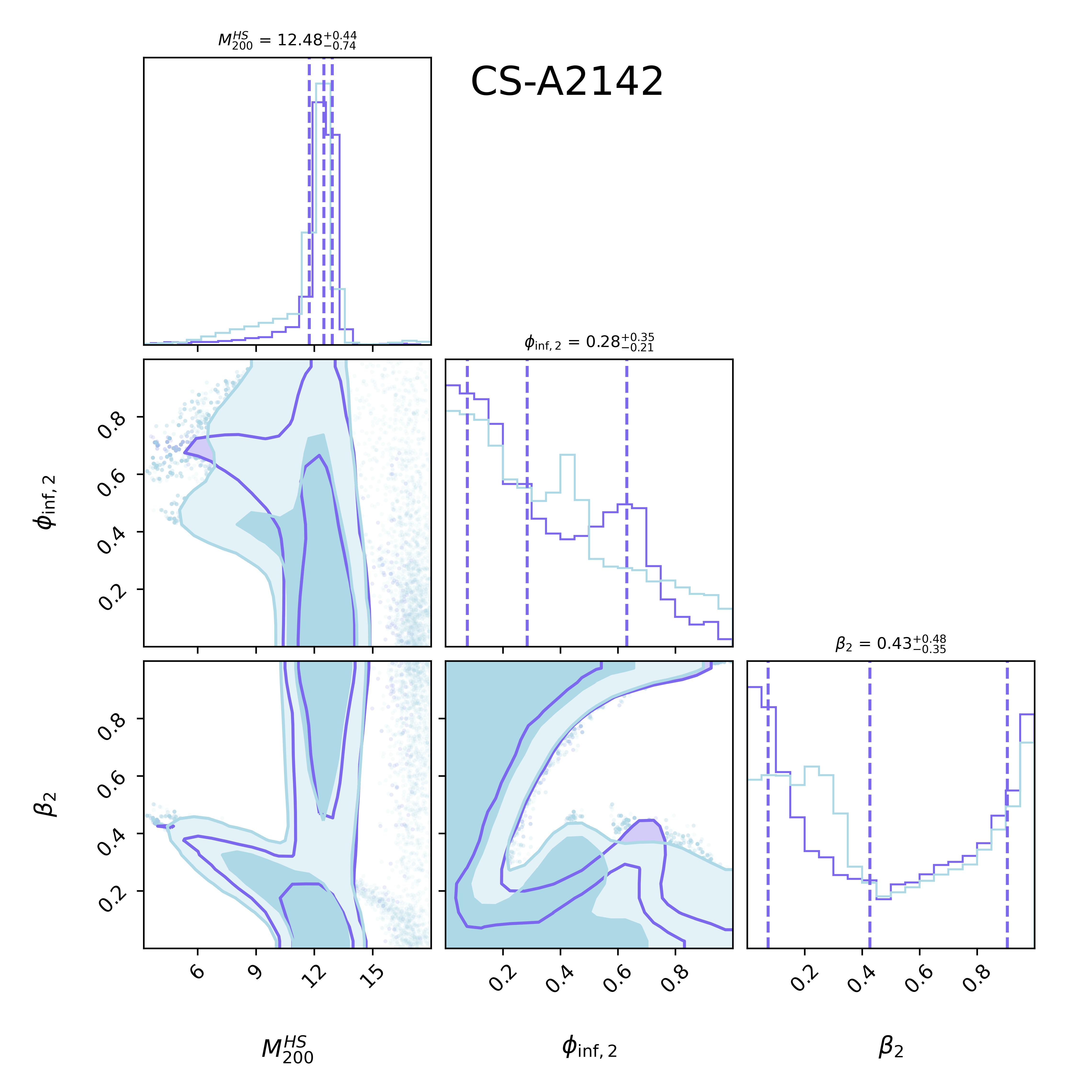}
    \caption{We show the $68\%$ and $95\%$ C.L. contours in the case of the \textit{Chameleon screening}, for the clusters A2029 (\textit{left}) and A2142 (\textit{right}). The {\textit{light blue}} filled contours show the constraints obtained using only the hydrostatic data. {Similarly, the contours outlined in dark blue} show the joint analysis including the caustic technique.}
    \label{fig:Contour_CS}
\end{figure*}

\begin{figure*}
    \centering
    \includegraphics[width=0.49\textwidth]{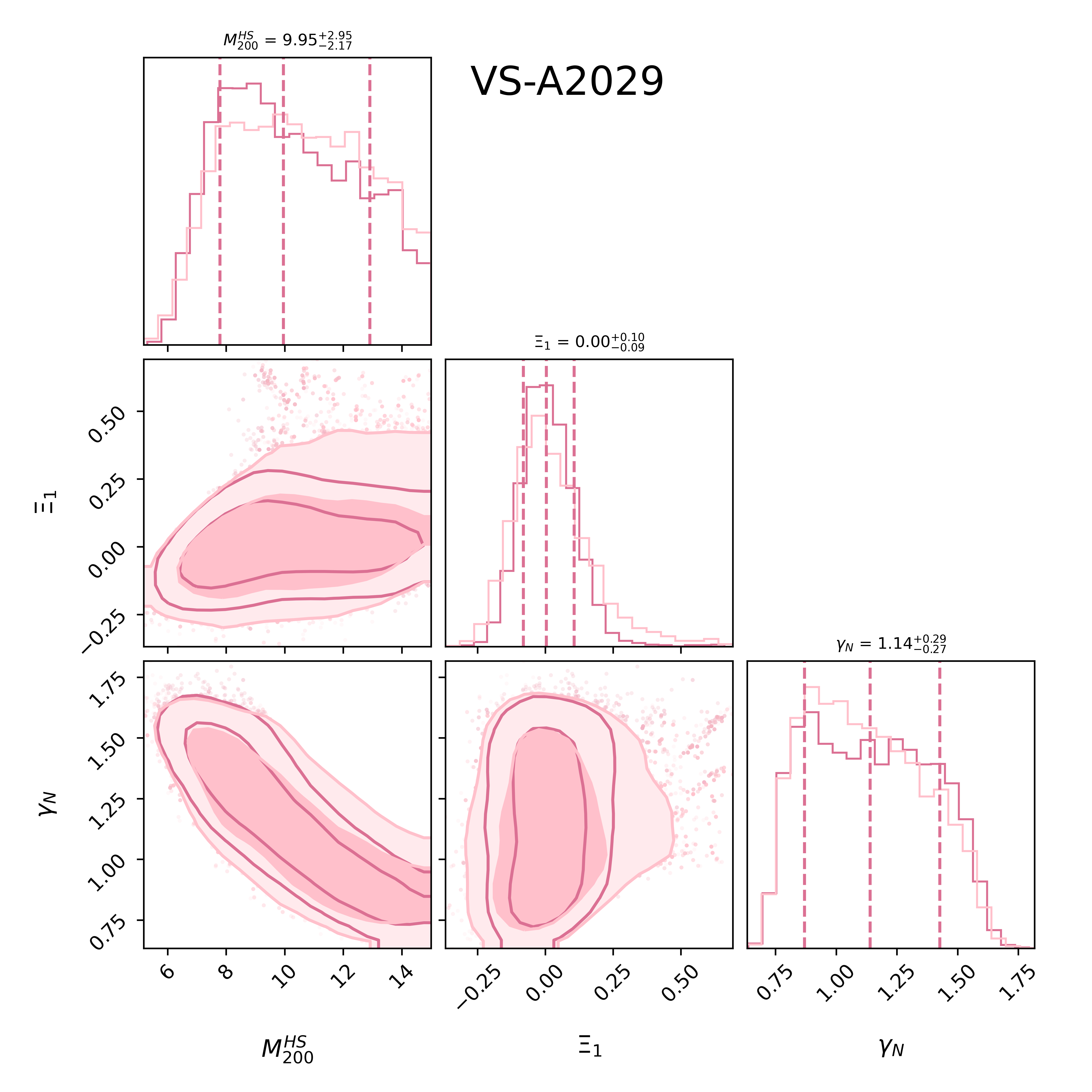}
    \includegraphics[width=0.49\textwidth]{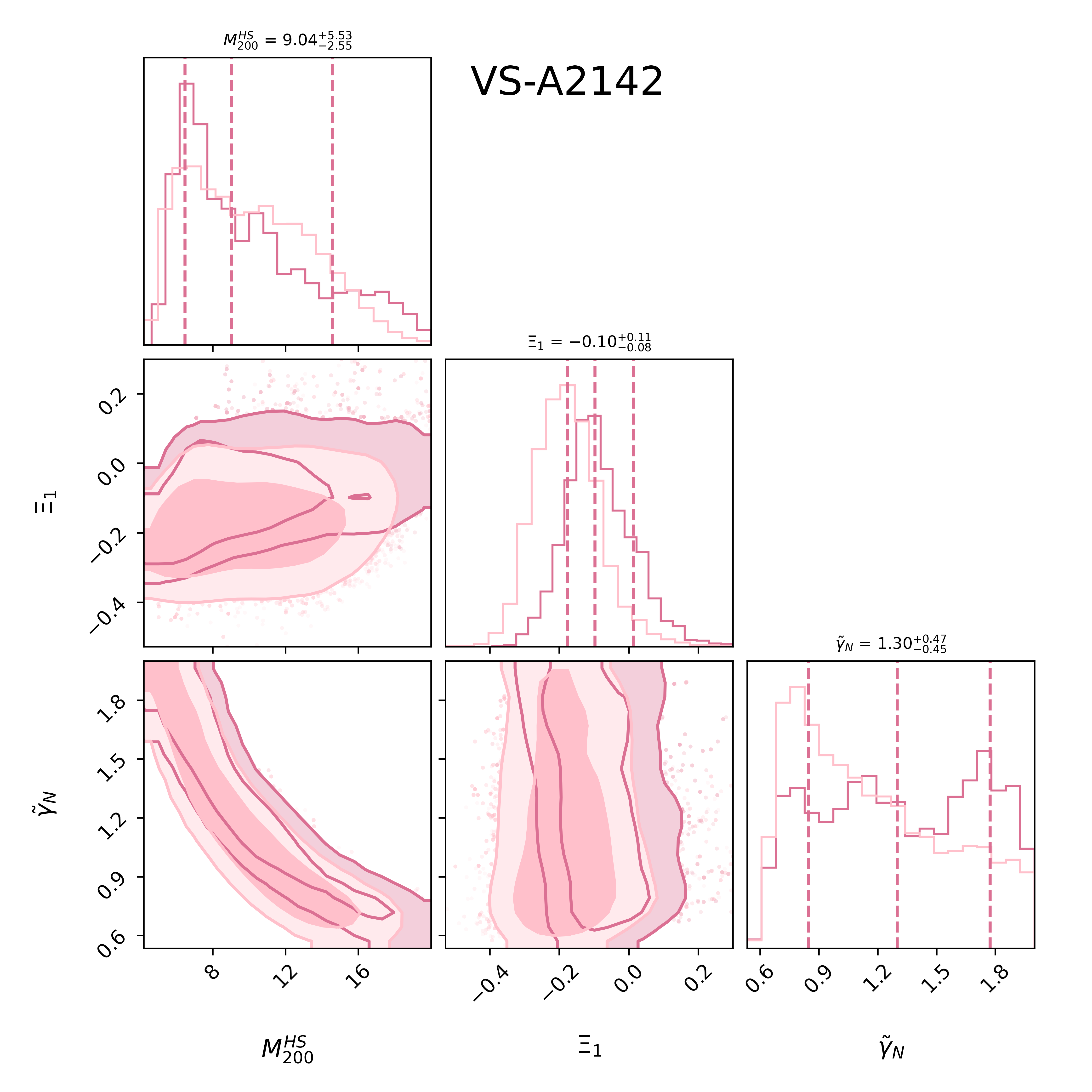}
    \caption{We show the $68\%$ and $95\%$ C.L. contours in the case of the \textit{Vainshtein screening}, for the clusters A2029 (\textit{left}) and A2142 (\textit{right}). The filled contours show the constraints obtained using only the hydrostatic data and {the contours outlined with darker red }show the joint analysis including the caustic technique.}
    \label{fig:Contour_VS}
\end{figure*}

As mentioned in the previous section, finding very good agreement between the hydrostatic and caustic masses inferred with our method in the GR case, now we extend to the modified gravity case simultaneously assessing both mass bias and the constraints on the additional fifth force parameters. In \Cref{tab:CS_Masses,tab:VS_Masses}, we compare the constraints on mass and modified gravity parameters. We show a simple comparison of the constraints when assuming $\fbr = 0.5$, to highlight the scenario wherein the caustic mass is inferred independent of the hydrostatic mass, and then a joint analysis is performed. This is a valid scenario given that the masses in the modified gravity models are less constrained and allow posteriors that can be consistent with $\fbr = 0.5$ and hence can provide varied constraints for the modified gravity parameters (see \Cref{sec:Incorrect_fbr}). Note that this comparison should be taken at face value only for illustrative purposes. In \Cref{fig:Contour_CS,fig:Contour_VS} we show the confidence level of the posteriors for the same, for the modified gravity parameters. 

{\renewcommand{\arraystretch}{1.6}%
    \setlength{\tabcolsep}{4pt}%

    \begin{table*}[!ht]
    \caption{{Same as \Cref{tab:GR_Masses}, but for the case of Chameleon screening. Note here we do not present the case of $\fbr = 0.65$. In columns 8-9 we present the constraints on the MG parameters, $\{\phitwo, \,\btwo\}$ obtained from our joint analysis.}}
    \label{tab:CS_Masses}

    \begin{tabular}{c c c c c c c c c c}
    \hline
        
        Cluster & \(\Rhs\)  & \(\Mhs \) & \(\fbr\) & \(\Rcau\)  & \(\Mcau \) & \(\Mjoint\) & $\phi_{\infty,2} $ & $\btwo$ & {\(\Mhs\)}/{\(\Mcau\)}\\
          & [Mpc] & \( [\Munit]\) & & [Mpc] & \( [\Munit]\) & \( [\Munit]\) & \(\beta = \sqrt{\frac{1}{6}}\)& \\
        
        \hline
        \hline
        
        \multirow{2}{*}{A2029} & \multirow{2}{*}{$2.04_{-0.13}^{+0.04}$}&
        \multirow{2}{*}  {$10.41_{-1.91}^{+0.62}$ }&
        HS&
        $2.08_{-0.10}^{+0.05}$&  $11.02_{-1.57}^{+0.75}$ & $10.84_{-0.58}^{+0.39} $ & $0.24_{-0.18}^{+0.29}$&
         -- & 
        $0.94_{-0.12}^{+0.13}$\\

        & & & 0.50 &  $1.86_{-0.03}^{+0.04}$ & $7.83_{-0.37}^{+0.38}$ & $7.87_{-0.46}^{+0.42}$ & $0.36_{-0.03}^{+0.12}$&
        $0.28_{-0.03}^{+0.03}$ &
        $1.32_{-0.23}^{+0.12}$ \\

        \multirow{2}{*}{A2142} & \multirow{2}{*}{$2.14_{-0.13}^{+0.03}$} &
        \multirow{2}{*}{$12.20_{-2.13}^{+0.53}$} & 
        HS & 
        $2.20_{-0.13}^{+0.07}$&
        $13.19_{-2.21}^{+1.20}$ & 
        $12.48_{-0.74}^{+0.44}$&
        $0.28_{-0.21}^{+0.35}$&
         -- &
        $0.91_{-0.10}^{+0.15}$\\

        & & & 0.50 &  
        {$2.02_{-0.04}^{+0.04}$}&
       $10.18_{-0.65}^{+0.59}$  & $10.28_{-0.64}^{+0.78}$ & $0.45_{-0.12}^{+0.36}$&
        $0.22_{-0.04}^{+0.04}$ &
        $1.18_{-0.19}^{+0.11}$\\


        \hline

    \end{tabular}
    \end{table*}
}

\textit{Chameleon screening}: As can be seen in the \Cref{fig:Contour_CS}, the joint analysis of the hydrostatic and caustic techniques helps reduce the degenerate region between mass ($\M$) and the coupling constant ($\btwo$) parameter, which we earlier elaborated upon in \cite{Boumechta:2023qhd}. Traditionally, the approach to alleviate this degeneracy and obtain constraints was to assume a mass prior obtained using weak lensing techniques and in \cite{Boumechta:2023qhd}, we introduced an internal mass prior that a prior of the mass obtained when restricting the analysis to $\btwo\geq 0.5$. In the current work, we utilize the Caustic technique to fulfill the role of a prior, however, the formalism is not to utilize a mass prior but to perform an importance sampling while simultaneously constraining the mass from the caustic technique. The ability of the Caustic technique to improve the constraints heavily relies on the mass estimates alone. In \Cref{tab:CS_Masses}, we show the joint mass estimates, which are once again tighter than either of the individual masses. In the last column of \Cref{tab:CS_Masses}, we show the mass bias between the hydrostatic and the caustic techniques, where we find a very good consistency within $\lesssim 1\sigma$. 

Needless to say, as can be seen in \Cref{fig:Contour_CS}, the constrain the $\phitwo$ for the case of $\btwo = \sqrt{1/6}$, which corresponds to the specific case of $\fofr$ gravity are also largely reduced for both the clusters. We infer a joint constraint of $\phitwo < 0.18$ at $99\%$ C.L., using only the two clusters here. This is a very good constraint and is comparable to the joint analysis present in \cite{Boumechta:2023qhd}, where five clusters were utilized with Weak lensing priors, $\phitwo<0.13$ at the same confidence level. This improvement here is essentially because the caustic mass we constrain taking into account the $\fbr$ already constrained using the hydrostatic data is much tighter than the WL masses present in \cite{Herbonnet:2019byy} (see Table I in \cite{Boumechta:2023qhd}), which were utilized in \cite{Boumechta:2023qhd} (see Table II therein).

\textit{Vainshtein screening}: In the case of Vainshtein screening, we have earlier established the constrained on the $\Xi$ parameter in \cite{Haridasu:2021hzq}. As can be seen in the third column of \Cref{tab:VS_Masses}, the current clusters in A2029, A2142 are very good examples for GR and mild deviation from GR case, respectively. In this context, the joint analysis is very helpful in constraining the modified gravity parameter $\Xi_1$. Firstly, as described in \Cref{sec:screening} (see also \cite{Haridasu:2021hzq}), within the Vainshtein screening the gravitational potential and the weak lensing potential are distinct a joint analysis would be needed to correctly constrain the running of the gravitational constant $\GNeff$. As the $\fbr$ necessary to constrain the caustic mass is the gravitational potential as well, we do not immediately break the degeneracy between the $\gNtilde$ and $\Mhs$. We present the constraints on the combination of these parameters in \Cref{tab:VS_Masses}. 

In the case of cluster A2029, we find a twice as tighter constraint for the parameter $\Xi_1 = 0.0 \pm 0.10$ with respect to the constraint of $\Xi_1 = -0.04^{+0.19}_{-0.12}$ obtained using only the hydrostatic data. This constraint is obtained while having no bias between the estimates, with $\Mhs/\Mcau = 0.98^{+0.15}_{-0.11}$, which shows a complete consistency. On the other hand, cluster A2142 which shows a mild $\sim 2\sigma$ deviation with $\Xi_1 \sim -0.20^{+0.10}_{-0.08}$ using only the hydrostatic data, is now more consistent with GR within $\sim 1\sigma$, having $\Xi_1 = -0.10^{+0.11}_{-0.06} $. Note that this is a shift in the parameters with no major improvement in the relative constraint. This improvement in the agreement with the GR is essentially because the masses estimated using the hydrostatic method and the caustic method have a bias of $\Mhs/ \Mcau = 0.81^{+0.09}_{-0.07}$, which is about the same significance $\geq 2\sigma$, as the earlier for modified gravity using the hydrostatic data alone. Nevertheless, we perform the joint analysis as the inconsistency is only of the order of $\sim 2 \sigma$. 

In this context, it is also interesting to note that for the cluster A2142, in the case of $\fbr = 0.5$, we find that the mass bias is no longer at a $\sim 2\sigma$ significance as earlier and is completely consistent with unity being $\Mhs/\Mcau = 1.05^{+0.13}_{-0.11}$. This is, however, accompanied by the joint constraint of $\Xi_1 = -0.21\pm 0.07$, which is a $3\sigma$ detection of modified gravity. We find this to be an interesting observation: if one demands the masses estimated using the two techniques to be consistent as a prior before assessing the joint constraint on the modified gravity parameters, this hypothetical case where the assumption of $\fbr = 0.5$ is not justified seems to provide better agreement and detection of deviation from GR. At face value, not having a mass bias, in this case, could imply that the extreme bias quoted in \cite{Ettori_2019} could be completely alleviated by simply changing the gravity model with the assumption of $\fbr$. However, it is important to assess the $\fbr$ correctly, before addressing the mass bias, as we have done consistently in our formalism even though the data shows a mild bias. As can be seen in \Cref{tab:VS_Masses}, this is not the case for cluster A2029, which is completely consistent with GR.

{\renewcommand{\arraystretch}{1.6}%
    \setlength{\tabcolsep}{4pt}%
    \begin{table*}[!ht]
    \caption{{Same as \Cref{tab:GR_Masses}, but for the case of Vainstein screening. Note here we do not present the case of $\fbr = 0.65$. In column 9, we show the constraints on the MG parameter, in this case, $\Xi_1$ obtained from our joint analysis.}}
    \label{tab:VS_Masses}

    \begin{tabular}{c c c c c c c c c c}
    \hline
        
        Cluster &  $R_{200}^{\mathrm{HS}}$ & 
        
        $\Xi_{1}$ &
        $\gNtilde\,M_{200}^{\mathrm{HS}}$&
        $\mathcal{F}_{\beta}(r)$ &
        $R_{200}^{\mathrm{Cau}}$ &
        $\gNtilde\,M_{200}^{\mathrm{Cau}}$ &
        $\gNtilde\,M_{200}^{\mathrm{Joint}}$ &
        $\Xi_{1}$ &
        {\(\Mhs\)}/{\(\Mcau\)}\\
        
        & [Mpc] & & & & [Mpc] & \( [\Munit]\) & \( [\Munit]\) &  & \\
        
        \hline
        \hline
        
        \multirow{2}{*}{A2029} & \multirow{2}{*}{$2.09_{-0.09}^{+0.11}$} & 
        \multirow{2}{*}{$-0.04_{-0.12}^{+0.19}$ } &
        \multirow{2}{*}{$11.16_{-1.40}^{+1.92}$ } &
        $\mathrm{HS}$ &
        $2.10_{-0.04}^{+0.04}$ & 
        $11.32_{-0.64}^{+0.68}$ &
        $11.15_{-1.01}^{+1.15}$ &  
        $-0.00_{-0.09}^{+0.10}$&
        $0.98_{-0.11}^{+0.15}$\\

       & & & & 0.50 &
       $1.86_{-0.03}^{+0.03}$ &  
       $7.84_{-0.38}^{+0.39}$ &
       $8.78_{-0.44}^{+0.53}$ & 
        $-0.18_{-0.06}^{+0.06}$&
        $1.43_{-0.19}^{+0.25}$\\

        \multirow{2}{*}{A2142} & 
        \multirow{2}{*}{$2.04_{-0.06}^{+0.07}$} & 
        \multirow{2}{*}{$-0.20_{-0.08}^{+0.10}$}&
        \multirow{2}{*}{$10.50_{-0.87}^{+1.04}$} &
        $\mathrm{HS}$ &
        $2.19_{-0.05}^{+0.05}$ & 
        $13.01_{-0.90}^{+0.96}$ &
        $11.80_{-0.94}^{+1.05}$ & 
        $-0.10_{-0.06}^{+0.11}$&
        $0.81_{-0.07}^{+0.09}$\\

        & & & & $0.50$ &  $2.01_{-0.04}^{+0.04}$ & 
        $10.04_{-0.62}^{+0.64}$&
        $10.21_{-0.58}^{+0.60}$ & $-0.21_{-0.06}^{+0.07}$&
        $1.05_{-0.11}^{+0.13}$\\


        \hline

    \end{tabular}
    \end{table*}
}

\section{Conclusion}
\label{sec:conclusion}

Joint analysis of different mass estimating techniques of galaxy clusters has been time and again utilized in constraining accurately the mass and consequently testing modified gravity theories \cite{Terukina:2013eqa, Wilcox:2015kna, Sakstein:2015aac, Haridasu:2021hzq, Pizzuti:2020tdl, Pizzuti:2022ynt, Saltas:2022ybg, Boumechta:2023qhd, Benetti:2023vxy}. In the same spirit, here we have assessed the ability of the caustic technique \cite{Diaferio_1997, Diaferio09, Serra_2010} to aid in testing modified gravity scenarios for the first time, to the best of our knowledge. 

The main implications and findings are as follows:

\begin{itemize}
    \item Firstly, we validate our formalism in GR, obtaining joint constraints on the mass of the two clusters. Essentially this is aided by the fact that we see no mass bias, which is quoted in \cite{Ettori:2018tus}. While we have already confirmed and found that the mass bias is reduced when the hydrostatic $\fbr$ is taken into account, another possible way to alleviate the mass bias would be by assuming a modified gravity scenario. 

    \item We have shown that the Caustic technique can robustly aid in testing the gravity within galaxy clusters. We have tested for the cases of Vainshtein and Chameleon screening. We also perform a hypothetical analysis, in case the caustic mass is estimated independently, how it could lead to spurious detection of modified gravity. 
    
    \item {In the case of the Vainshtein screening we were able to obtain tighter constraints on the free parameter with respect to the solely hydrostatic analysis. In particular, for the cluster A2142, the X-ray data provided a constraint below the expectation of GR, $\Xi_1 \sim -0.2$, with $\sim2\sigma$ significance, while the joint analysis reduces the tension to $\Xi_1 = -0.10^{+0.11}_{-0.06} $ at $\lesssim 1\sigma $. For cluster A2029, the consistency with GR remains intact with an improvement on the bounds at the same confidence level of a factor of $\sim 2$ in the joint analysis.} 

    \item For the case of Chameleon screening we find that the caustic technique aids in reducing the degeneracy between the mass estimates ($\M$) and the coupling constant ($\btwo$), as can be seen in \Cref{fig:Contour_CS}. Given this we estimate a competitive joint constraint of $\phitwo \lesssim 0.18$ at $3\sigma$ C.L., using only the two clusters utilized in this work. This further reinforces the utility of the caustic technique in constraining the modified gravity parameters.

\end{itemize}

Although in the current work, we have limited ourselves to two clusters as proof of concept to explore the possibility, the formalism can be easily extended to several individual clusters stacked \cite{Gifford:2016plw, Svensmark:2014cba, Pizzardo:2023idp, Pizzardo:2023wdo} for both the hydrostatic data (see for example \cite{Wilcox:2015kna, Wilcox:2016guw}) and caustic phase space, which we intend to extend as one of the future directions. {Note that our constraints obtained using only individual clusters are already an improvement (see \Cref{sec:MG_mass}) over stacked analysis in \cite{Wilcox:2015kna}, which utilizes priors from weak lensing in contrast to our approach of using the caustic technique. This improvement is mainly due to two factors: improvement in the hydrostatic data \cite{Ettori_2019}, as reported in our earlier work \cite{Boumechta:2023qhd} and the joint analysis with caustic information performed here. However, the caustics allow us to constrain better the overall posteriors of $\{\phitwo,\, \btwo\}$, reducing the multimodal behavior for the $\phitwo$ constrains for $\beta = \sqrt{1/6}$ in the case of $\fofr$ model. }

We anticipate the current formalism to help perform a thorough investigation of the modified gravity models utilizing the caustics as an independent observable alongside the already well-explored weak lensing and the hydrostatic datasets. we also intend to extend our formalism to other modified gravity, non-standard dark matter scenarios \cite{Gandolfi:2023hwx, Benetti:2023vxy}, with non-local effects on the dark matter mass profiles, in future communications. Aside from these we also intend to joint analysis of the kinematics, dynamics, and weak lensing datasets exploring the potential to constrain the modified gravity parameters and break degeneracies within them. 

\section*{Acknowledgements}
The authors are grateful to Antonaldo Diaferio for insightful discussions and valuable comments on the draft. AL has been supported by the EU H2020-MSCA-ITN-2019 Project 860744 `BiD4BESt: Big Data applications for Black Hole Evolution Studies' and by the PRIN MIUR 2017 prot. 20173ML3WW, `Opening the ALMA window on the cosmic evolution of gas, stars, and supermassive black holes'. BSH is supported by the INFN INDARK grant and acknowledges support from the COSMOS project of the Italian Space Agency (cosmosnet.it). CB acknowledges support from the COSMOS project of the Italian Space Agency (cosmosnet.it), and the INDARK Initiative of the INFN (web.infn.it/CSN4/IS/Linea5/InDark).

\bibliography{bibliography} 

\appendix

\section{Error estimation in the caustic technique}
\label{sec:caustic_error}
In this section, we highlight a few systematics that affect the estimated caustic mass. The caustic technique heavily relies on the methods utilized to set the so-called `Caustic' surface. The caustic surface is estimated as the edge of the galaxy cluster assessed through the distribution of the galaxies in the line-of-sight velocity and the projected radius from the center of the clusters. We utilize our importance sampling-like approach instead of a likelihood-based technique as the error estimation of the caustic surface relies on the density of the galaxies, which can be estimated at every radius instead of data with error bars that can be incorporated into a simple Gaussian likelihood. 

\section{Comment on the anisotropy profile $g_{\beta}(r)$ assumption}
\label{sec:Anisotropy_prof}

As mentioned in the main text, we have assumed a fixed anisotropy profile of the galaxy phase space in the current analysis. To check the robustness of our statement, we also fit the anisotropy profile using the \textsc{MG-MAMPOSSt} code \cite{Pizzuti:2020tdl}, which provides joint mass density and orbit anisotropy reconstructions in GR and several modified gravity scenarios. More in detail, \textsc{MG-MAMPOSSt} relies on data in the projected phase space of member galaxies; it solves the spherically-symmetric Jeans' equation, assuming a shape for the three-dimensional velocity distribution and dynamical relaxation of the system. The code implements a parametric determination of the kinematic quantities (i.e. gravitational potential, number density of galaxies, and velocity anisotropy) by performing a Maximum Likelihood fit in the projected phase space.

Assuming both GR and one of the modified gravity frameworks, namely Chameleon screening, we reconstruct the mass distribution and the anisotropy profile of A2029 and A2142. In all cases, we adopt an NFW model for the matter density distribution, while the velocity anisotropy is parametrized with a generalized Tiret model (e.g. \cite{Mamon19}),

\begin{equation}
\label{eq:betagT}
\beta_\text{gT}(r)=\beta_0 + (\beta_\infty + \beta_0)\frac{r}{r-r_\beta}\,
\end{equation}

where $\beta_0$, $\beta_\infty $ represent the central anisotropy and the anisotropy at a large distance from the center, respectively; $r_\beta$ is a characteristic radius which we assume to be equivalent to the scale radius $r_\text{s}$ of the total mass profile. \Cref{eq:betagT} provides a quite general description of possible orbit anisotropy profiles in galaxy clusters. For both clusters, we consider galaxies that lie in projection within $R\sim 1.1\, R_{200}^\text{HS}$, to ensure the validity of Jean's equation. Nevertheless, we checked that reasonable changes ($10\%$) in that limit provide negligible effects on the final results. We further exclude the central region $R<0.05\, \Mpc$, where the dynamics are dominated by the Brighter Central Galaxy.

We use than the anisotropy profile reconstructed with \textsc{MG-MAMPOSSt} to validate the statement that our assumption has a negligible effect on the inferred final caustic mass. Essentially, this is because the anisotropy profile, which requires the knowledge of $R_{200}$,  is degenerate with the cluster viral radius, and hence with the total mass. However, since the hydrostatic technique constraints the same quantity much more tightly than the anisotropy profile itself, we find that fixing it to a constant is largely valid and makes minimal changes to the conclusions made here.

\section{Incorrect assumptions of $\fbr$ in modified gravity}
\label{sec:Incorrect_fbr}
\begin{figure*}
    \centering
    \includegraphics[width=0.47\textwidth]{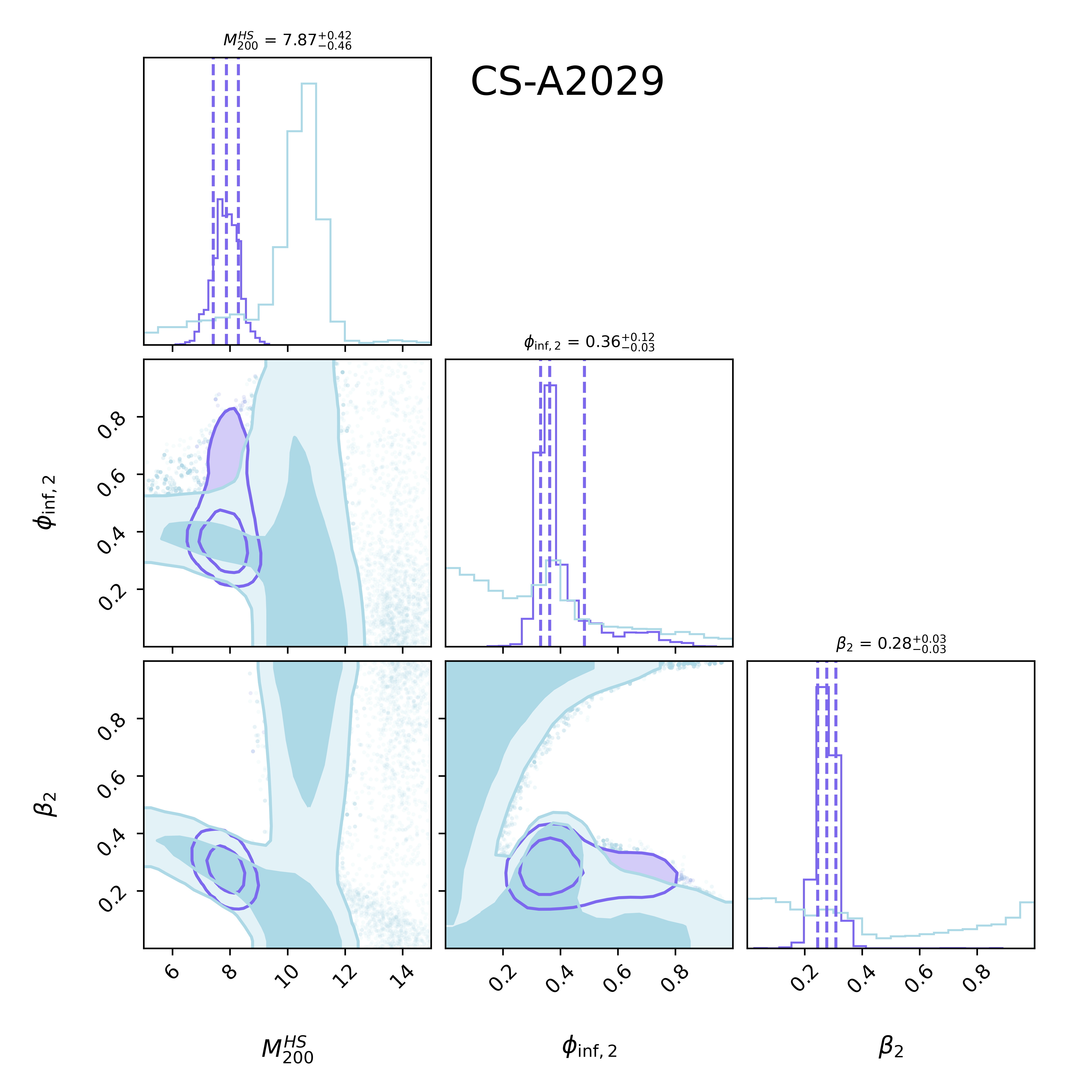}
    \includegraphics[width=0.47\textwidth]{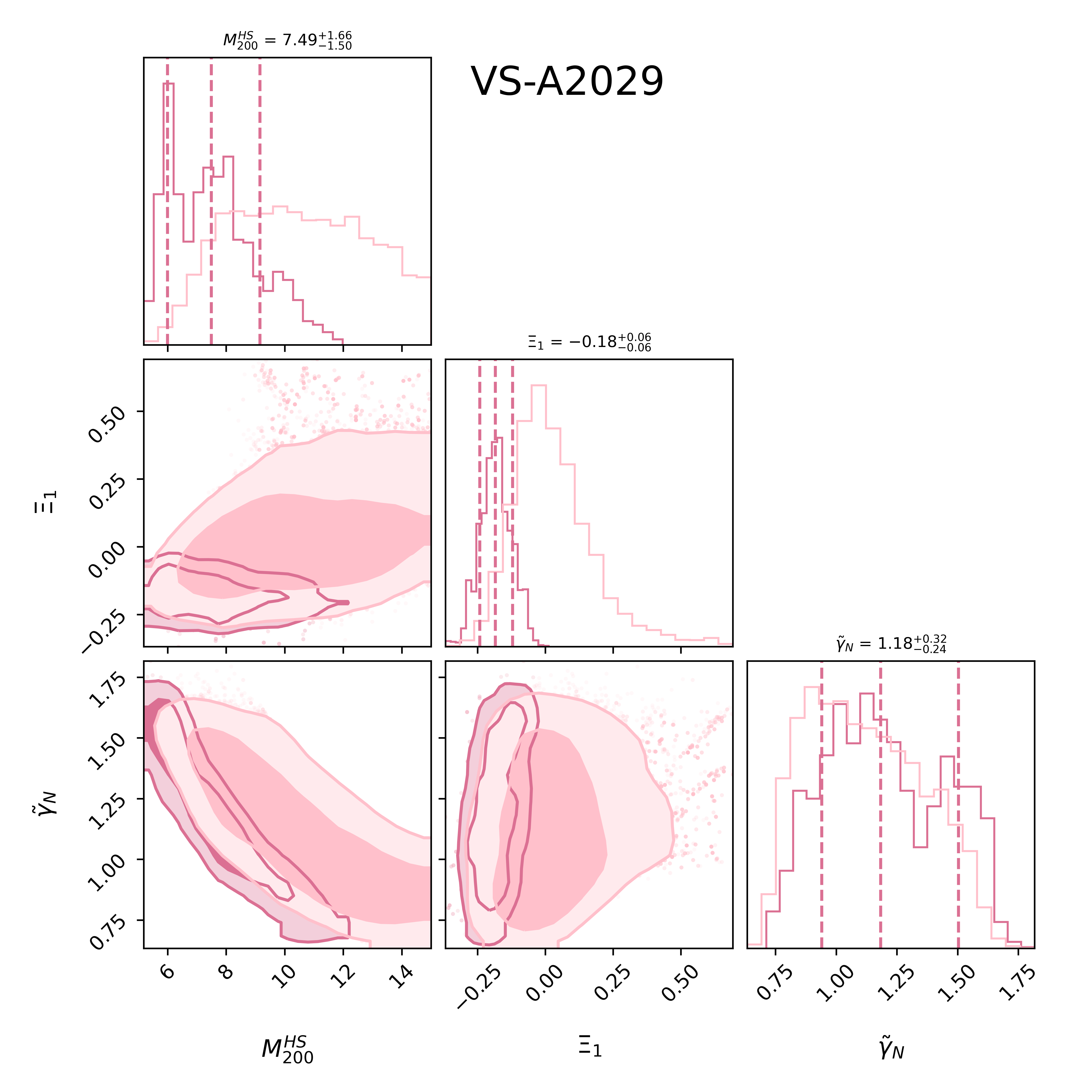}
    \caption{\textit{Left:} We show the constraints on the modified gravity parameters in the case of the chameleon screening when assuming a $\fbr = 0.5$. \textit{Right:} Same as \textit{Left} panel, but for the case of the Vainshtein screening. }
    \label{fig:A2029_CS_VS_const}
\end{figure*}

It is well established that the $\fbr$ in the inner regions of the cluster is not a constant and assuming so can lead to biases. In this section, we show as an example, the potential of the caustic technique to very well constrain the modified gravity scenarios in case the $\fbr$ derived from hydrostatic equilibrium differs from that constrained using alternate techniques or simulations. This is particularly important in the modified gravity scenarios, where the constraints on the mass are relaxed when using the hydrostatic equilibrium data (see \Cref{fig:mass_profiles}). As a test case, we show this through the example of the cluster A2029, when assuming $\fbr = 0.5$ as an independent estimate. When performing the joint analysis the essential improvement in the constraints on the modified gravity model parameters is valid if and only if the caustic masses are inferred utilizing the apriori constrained masses using the hydrostatic technique. For instance, in \Cref{fig:A2029_CS_VS_const}, we show the joint constraints from the importance sampling against the hydrostatic-only constraints, wherein the caustic technique assumes a constant $\fbr= 0.5$. One can immediately notice the well-bounded constraints on the parameters $\{\phitwo, \btwo\}$, which are spurious posteriors indicating evidence for modified gravity, in the current comparison. However, as shown in the \Cref{fig:Contour_CS,fig:Contour_VS}, for the cluster A2029 while the caustic technique helps to tighten the posterior assuming the correct $\fbr$, does not indicate any deviation from GR. Needless to say, we make equivalent conclusions also in the case of the cluster A2142.

\section{Mass profiles}
\label{sec:mass_profiles}
\begin{figure*}
    \centering
    \includegraphics[scale=0.28]{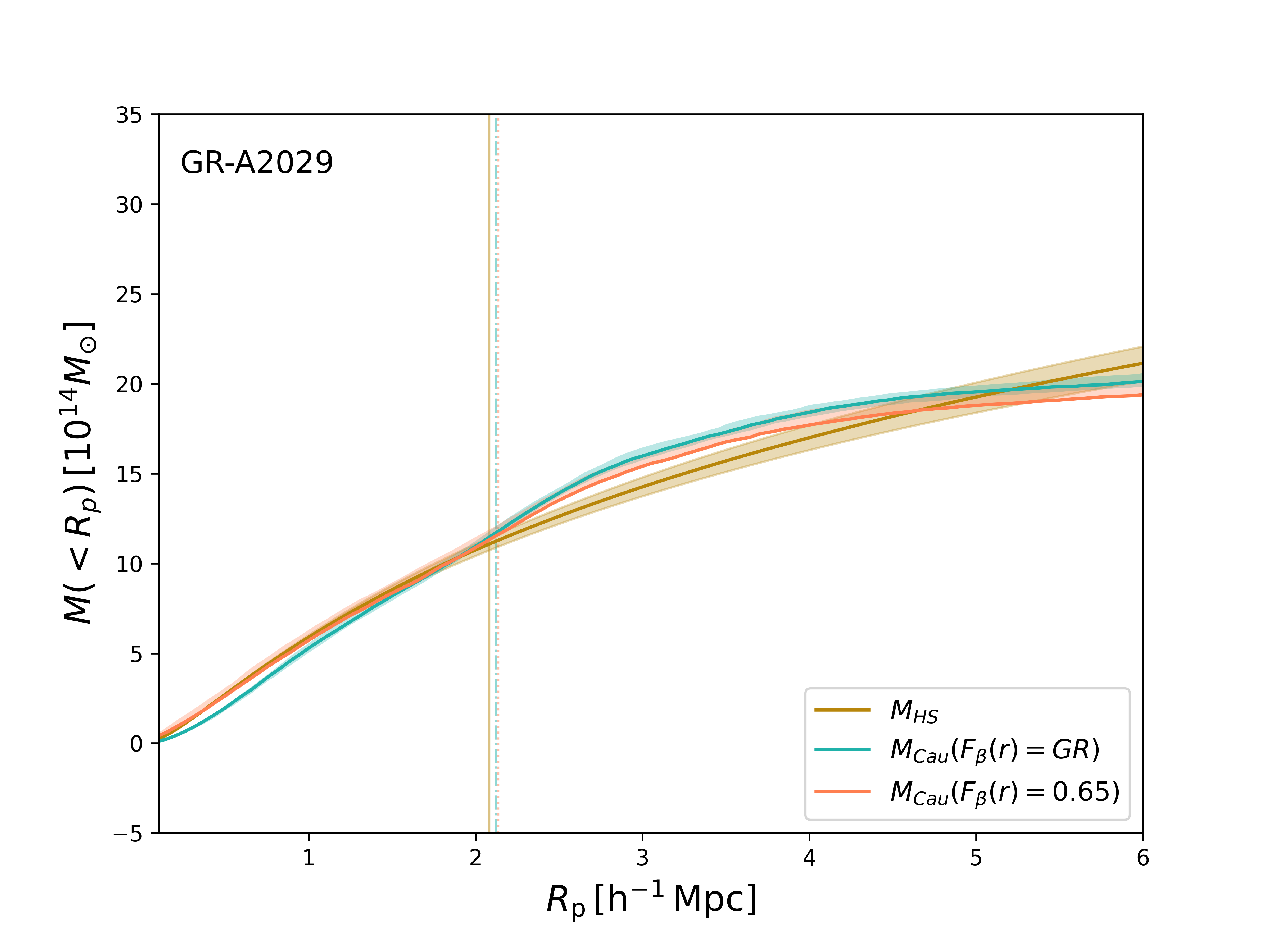}
    \includegraphics[scale=0.28]{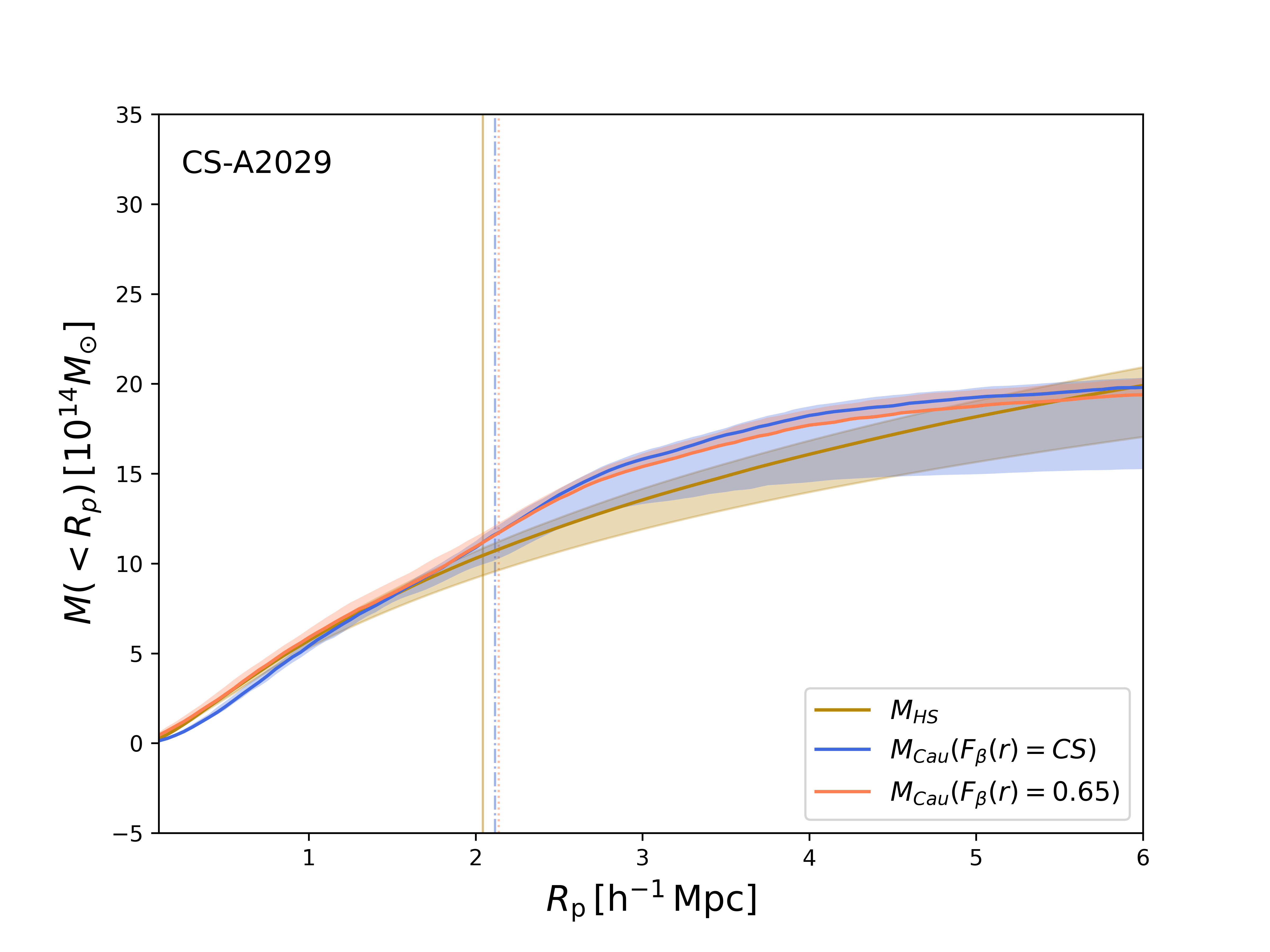}
    \includegraphics[scale=0.28]{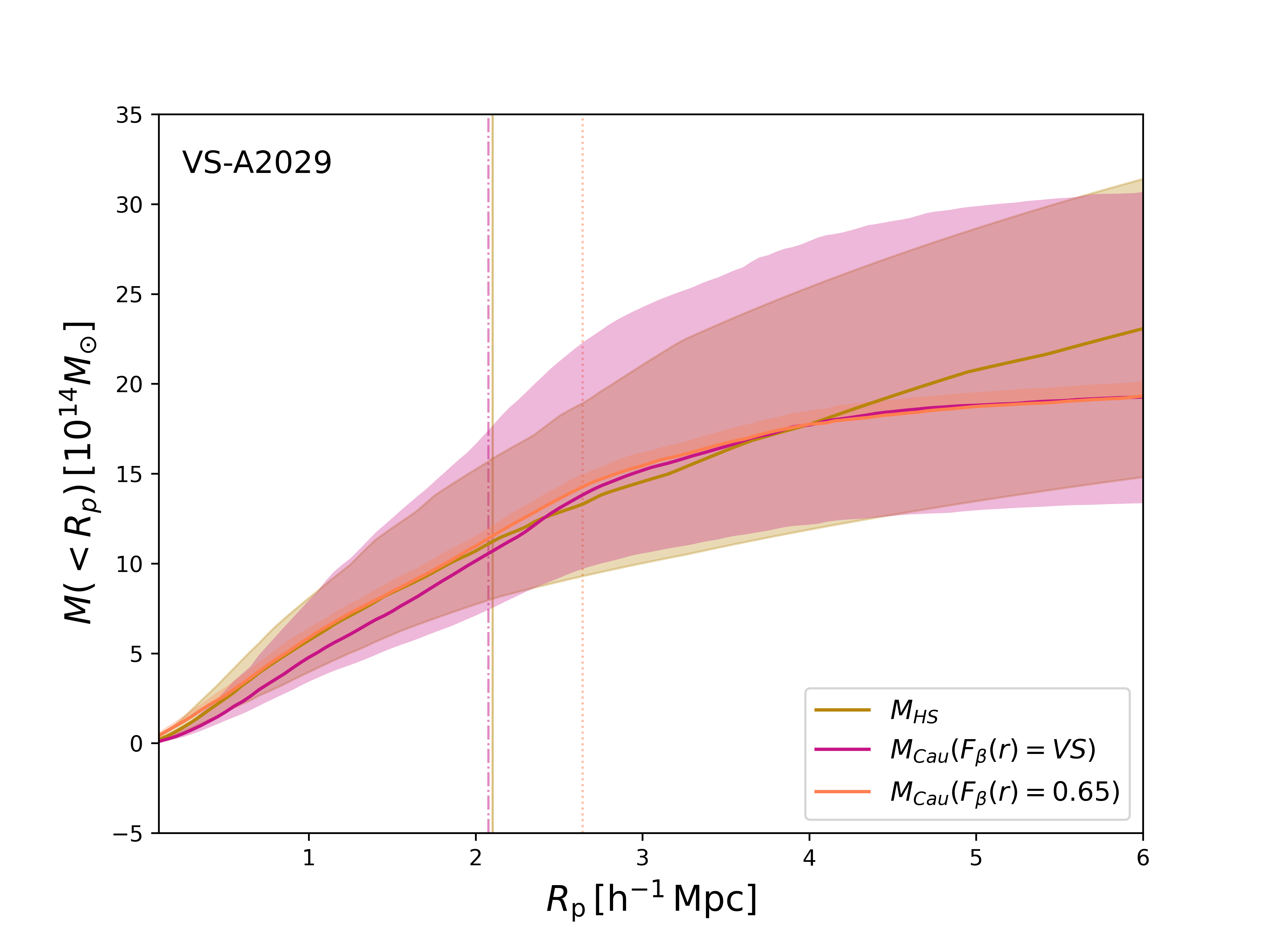}

    \centering
    \includegraphics[scale=0.28]{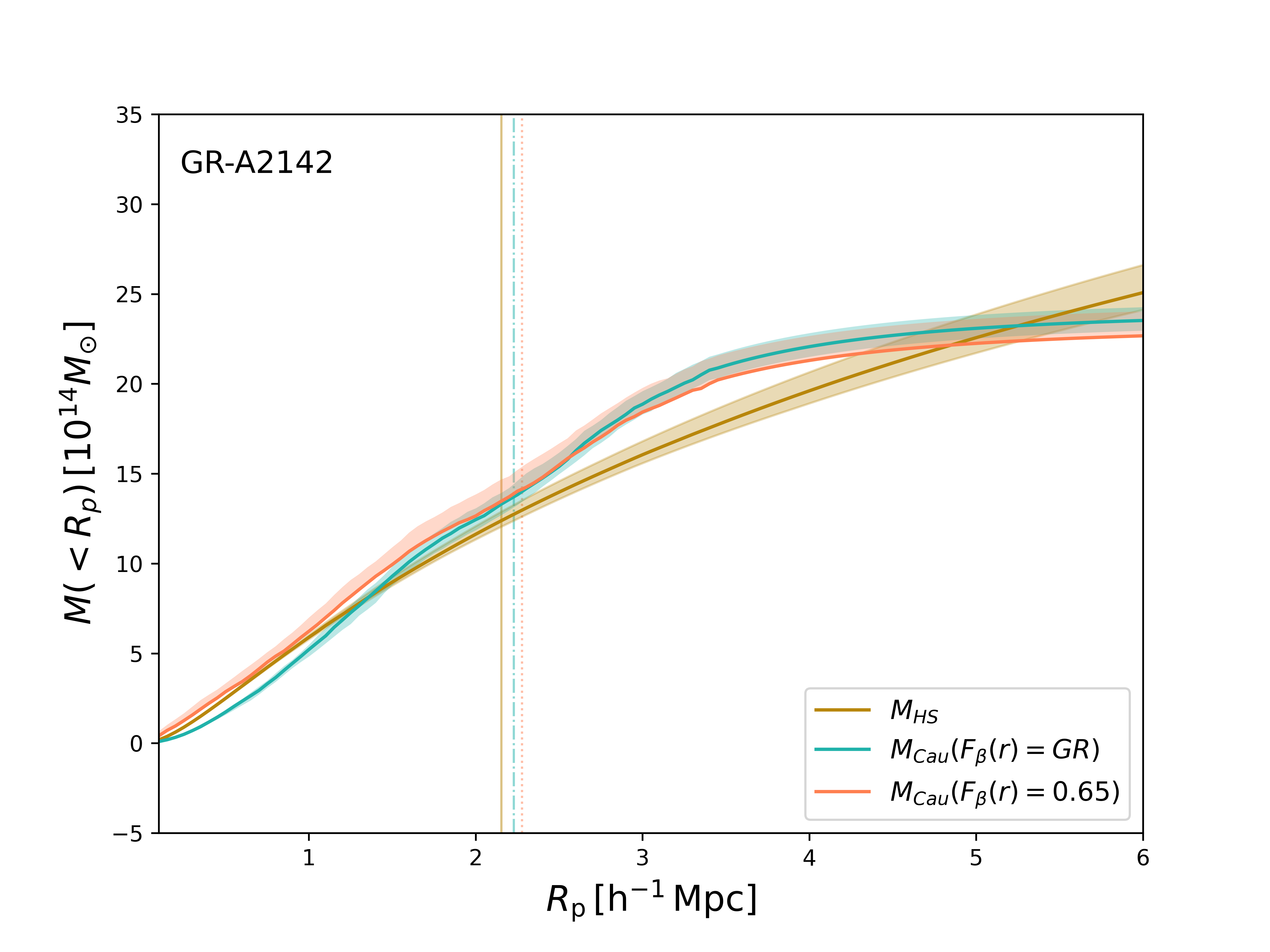}
    \includegraphics[scale=0.28]{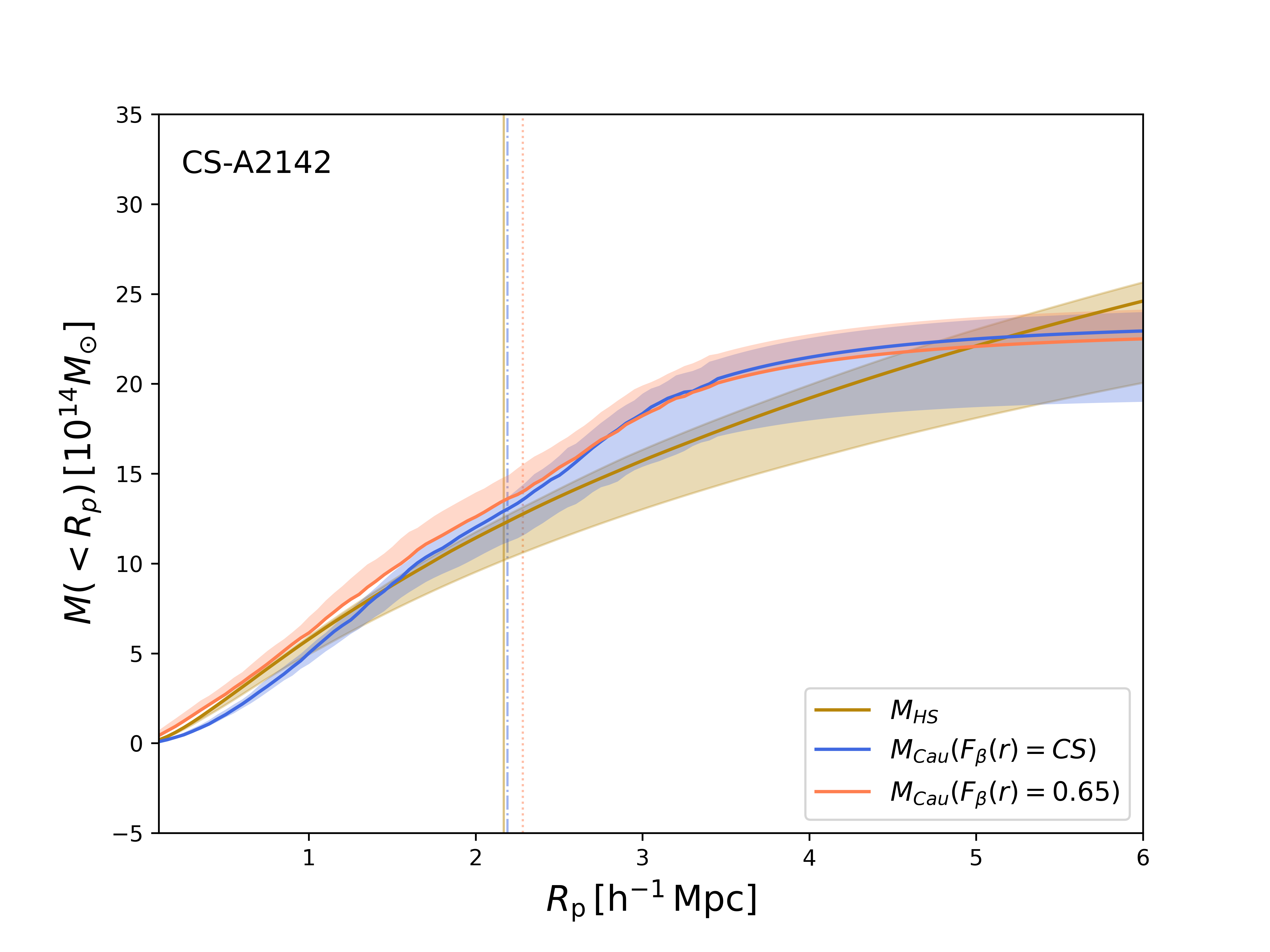}
    \includegraphics[scale=0.28]{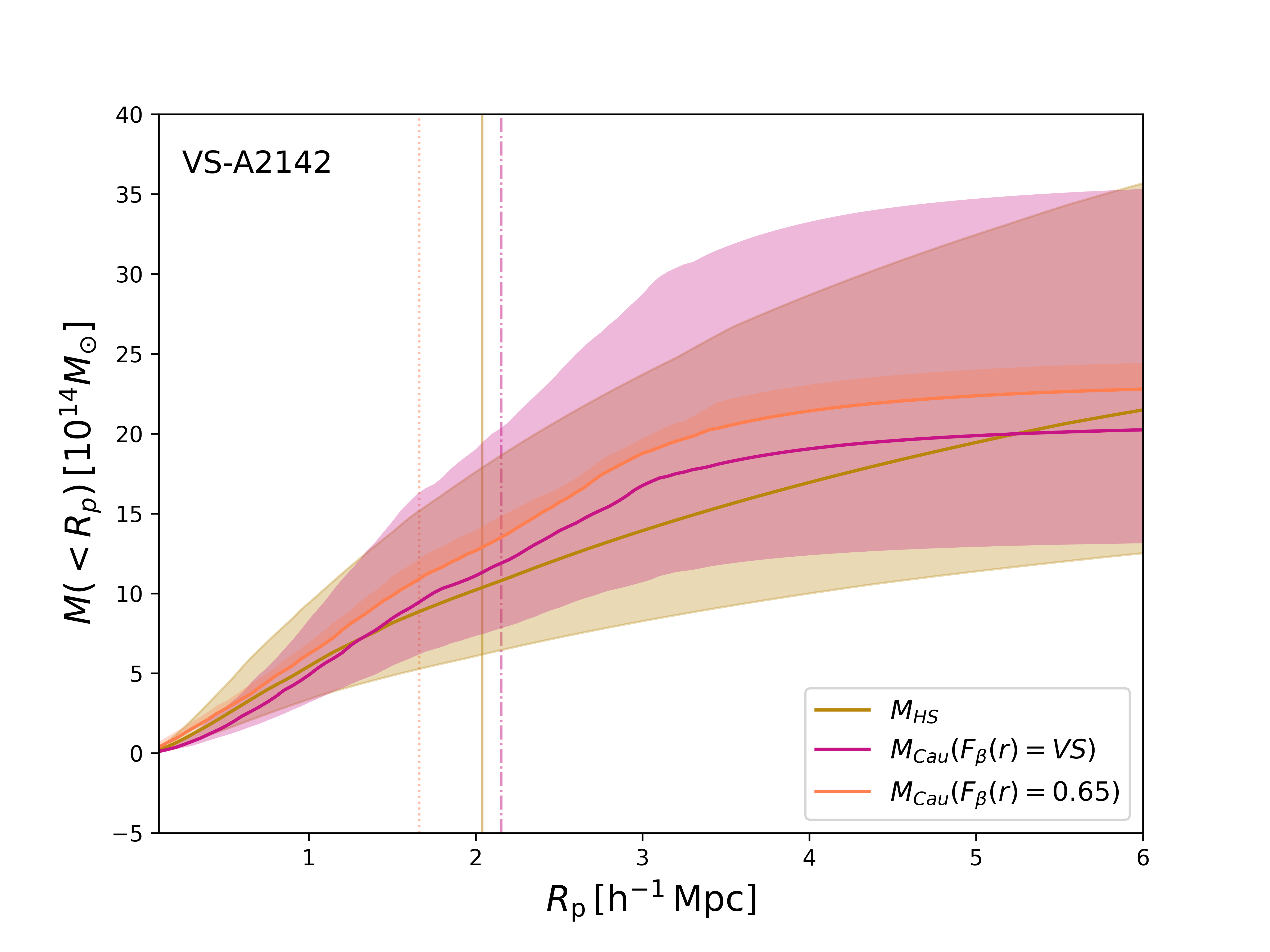}
    \caption{We compare the mean and the $68\%$ confidence level regions of the mass profiles obtained using the hydrostatics and caustics techniques. \textit{Top}: Mass profiles in the case of GR (\textit{left}), chameleon screening (\textit{centre}) and Vainshtein screening (\textit{right}). In each of the panels, we show the NFW-based mass profile (yellow), model-independent reconstructed mass profile using the caustic technique, assuming hydrostatic $\fbr$ (green) and constant $\fbr = 0.65$ (red). \textit{Bottom}: Same as \textit{Top} panel, for the cluster A2142.}
    \label{fig:mass_profiles}
\end{figure*}

In \Cref{fig:mass_profiles}, we show the mass profiles in various scenarios. We find that the model-independent caustic mass profile is in very good agreement with the NFW-based mass profile constrained using the hydrostatic data. We also find an overall slightly lower mass profile when assuming the $\fbr = 0.5$ and a very good agreement when using $\fbr = 0.65$, while showing no discernable change in the shape of the profile. The caustic mass profile also shows a tendency to get flattened in the outskirts of the cluster, which is a better assessment of the mass profile in contrast to assuming a mass profile such as NFW which asymptotically increases. While we retain the comparison in the main text to the mass $\M$ of the cluster, which is convenient for performing statistical analysis, we could also place upper limits on the total mass of the cluster using the caustic mass profile. Note that the mass profiles reconstructed using the caustic technique flatten out, seemingly suggesting an asymptotically converging total mass of the cluster. However, this is only an artifact of the reconstruction technique, wherein the caustic surface, $\mathcal{A}^{2}(r)$ artificially goes to zero where no galaxies are present. This in turn implies the integrand in \Cref{eqn:Mass_Profile} goes to zero providing constant mass values after a certain radial range. Therefore, we note that while we show the mass profiles going out to 6 Mpc, the nature of the increase of mass should be inferred from the NFW-based extrapolation or the flattened caustic mass profiles. In this context, it is also possible to utilize a value of mass say at $\Delta = 100$ to perform our joint analysis, which we intend to leave for a future investigation while improving the joint analysis.

\end{document}